\documentclass[useAMS,usenatbib]{mn2e}
\usepackage{psfig}
\usepackage[a4paper]{hyperref}
\usepackage{amssymb}
\usepackage{aas_macros} 

\newcommand{\rmsub}[2]{#1_{\rm #2}} 

\newcommand{\Teff}{\mbox{$T_{\mbox{\scriptsize eff}}\,$}}

\title[Efficient transit candidate identification]{Efficient identification of exoplanetary transit candidates from SuperWASP light curves
}
\author[A. Collier Cameron et al]
{
A. Collier Cameron$^{1}$\thanks{E-mail:acc4@st-and.ac.uk},
D.M. Wilson $^{2}$,
R.G. West $^{3}$,
L. Hebb$^{1}$,
X.-B. Wang$^{4,1}$,
\newauthor
S. Aigrain$^{5}$,
F. Bouchy$^{6}$,
D.J. Christian$^{7}$,
W.I. Clarkson$^{8}$,
B. Enoch$^{9}$,
M. Esposito$^{10}$,
\newauthor
E. Guenther$^{10}$,
C.A. Haswell$^{9}$,
G. H\'ebrard$^{6}$,
C. Hellier$^{2}$,
K. Horne$^{1}$,
J. Irwin$^{11}$,
\newauthor
S.R. Kane$^{12}$,
B. Loeillet$^{13}$,
T.A. Lister$^{2,14}$,
P. Maxted$^{2}$,
M. Mayor$^{15}$,
C. Moutou$^{13}$,
\newauthor
N. Parley$^{9}$,
D. Pollacco$^{7}$,
F. Pont $^{15}$,
D. Queloz$^{15}$,
R. Ryans$^{7}$,
I. Skillen$^{16}$,
\newauthor
R. A. Street$^{7,14/17}$,
S. Udry$^{15}$,
and
P.J. Wheatley $^{18}$
\\
\\
$^{1}$School of Physics and Astronomy, University of St Andrews, North Haugh, St Andrews, Fife KY16 9SS, UK.\\
$^{2}$Astrophysics Group, Keele University, Staffordshire, ST5 5BG.\\
$^{3}$Department of Physics and Astronomy, University of Leicester, Leicester, LE1 7RH, UK.\\
$^{4}$National Astronomical Observatories/Yunnan Observatory, Chinese Academy of Sciences, Kunming, China.\\
$^{5}$School of Physics, University of Exeter, Stocker Road, Exeter, EX4 4QL, UK.\\
$^{6}$Institut d'Astrophysique de Paris, CNRS (UMR 7095) --  Universit\'e Pierre \&\ Marie Curie, 98$^{bis}$ bvd. Arago, 75014 Paris, France.\\
$^{7}$ARC, Main Physics Building, School of Mathematics \&\ Physics, Queen's University, University Road, Belfast, BT7 1NN, UK.\\
$^{8}$STScI, 3700 San Martin Drive, Baltimore, MD 21218, USA.\\
$^{9}$Department of Physics and Astronomy, The Open University, Milton Keynes, MK7 6AA, UK.\\
$^{10}$Th{\"u}ringer Landessternwarte Tautenburg, Karl-Schwarzschild-Observatorium, D-07778 Tautenburg, Germany.\\
$^{11}$Institute of Astronomy, University of Cambridge, Madingley Road, Cambridge, CB3 0HA, UK.\\
$^{12}$Department of Astronomy, University of Florida, 211 Bryant Space Science Center, Gainesville, FL 32611-2055, USA.\\
$^{13}$Laboratoire d'Astrophysique de Marseille, BP 8, 13376 Marseille Cedex 12, France.\\
$^{14}$Las Cumbres Observatory, 6740B Cortona Drive, Goleta, CA 93117, USA.\\
$^{15}$Observatoire de Gen\`eve, Universit\'e de Gen\`eve, 51 Ch. des Maillettes, 1290 Sauverny, Switzerland.\\
$^{16}$Isaac Newton Group of Telescopes, Apartado de Correos 321, E-38700 Santa Cruz de la Palma, Tenerife, Spain. \\
$^{17}$Dept. of Physics, Broida Hall, University of California, Santa
Barbara, CA 93106-9530, USA.\\
$^{18}$Department of Physics, University of Warwick, Coventry CV4 7AL, UK.\\
}
\begin{document}

\date{Accepted 2007 July 2. Received 2007 June 29; in original form 2007 May 18}

\pagerange{\pageref{firstpage}--\pageref{lastpage}} \pubyear{2007}

\maketitle

\label{firstpage}

\begin{abstract}
Transiting extrasolar planets constitute only a small fraction of the  range of stellar systems found to display periodic, shallow dimmings in wide-field surveys employing small-aperture camera arrays. Here we present an efficient selection strategy for follow-up observations, derived from analysis of the light curves of a sample of 67 SuperWASP targets that passed the selection tests we used in earlier papers, but which have subsequently been identified either as planet hosts or as astrophysical false positives. We determine the system parameters using Markov-chain Monte Carlo analysis of the SuperWASP light curves. We use a constrained optimisation of $\chi^2$ combined with a Bayesian prior based on the main-sequence mass and radius expected from the 2MASS $J-H$ colour. The Bayesian nature of the analysis allows us to quantify both the departure of the host star from the main-sequence mass-radius relation and the probability that the companion radius is less than 1.5 Jupiter radii. When augmented by direct light curve analyses that detect binaries with unequal  primary and secondary eclipses, and objects with aperture blends that are resolved by SuperWASP, we find that only 13 of the original 67 stars, including the three known planets in the sample, would qualify for follow-up. This suggests that planet discovery ``hit rates" better than one-in-five should be achievable. In addition, the stellar binaries that qualify are likely to have astrophysically interesting stellar or sub-stellar secondaries.
\end{abstract}


\section{Introduction}

Wide-field photometric surveys using commercial, small-aperture camera lenses and large-format CCDs have yielded several discoveries of new transiting extrasolar planets \citep{alonso2004,mccullough2006,odonovan2006,cameron2007,bakos2007a,bakos2007b,burke2007,odonovan2007}. Deeper surveys based on light curves from the Optical Gravitational Lensing Experiment (OGLE) project have produced a further five such planets
 (\citealt{konacki2003}; \citealt{bouchy2004}, \citealt{konacki2004}; \citealt{pont2004}; \citealt{konacki2005}). 

The most efficient followup strategies (in terms of telescope time expended per false positive) are high-resolution radial velocity surveys using 2-m and larger telescopes.  \citet{pont2005a} points out that a single spectrum quickly unmasks double-lined spectroscopic binaries and rapid rotation resulting from tidal locking of the primary's rotation by a stellar-mass companion. It also reveals pressure-broadening of strong absorption-line wings,  distinguishing K giants from K dwarfs. Second-epoch spectra then eliminate the narrow, single-lined stellar spectroscopic binaries among the surviving main-sequence stars. All the teams engaged in such searches report high rates of astrophysical false positives. For example, \citet{pont2005a} carried out a detailed analysis of targets followed up in the OGLE Carina field. They found that nearby blended eclipsing binaries, grazing equal-mass stellar binaries and transits by planet-sized stars near the bottom of the main sequence are the most common impostors.  

Several authors have described methods for pre-selecting targets for such programmes in order to minimise the number of astrophysical false positives in samples selected for radial-velocity followup. \citet{seager2003} pointed out that the relative radii of the planet and star can be estimated from the transit depth, while the durations of ingress and egress can break the degeneracy between the impact parameter and the star's radius. The transit duration then yields the radius of the primary, and Kepler's third law establishes its density, allowing dwarf-giant separation. The masses of bona-fide main-sequence primaries are then estimated from the mass-radius relation.

In practice, however, the signal-to-noise ratios of transit profiles secured with wide-field cameras are seldom good enough to determine the duration of ingress and egress reliably. This problem led \citet{tingley2005} to propose a simpler test, based on transit duration, depth and orbital period, for the selection of transit candidates worthy of followup. \citet{drake2003} and \citet{sirko2003} used ellipsoidal variations in the out-of-transit light curve to eliminate stellar binaries. Even with careful pre-selection on transit depth, duration, and primary-star colour and ellipsoidal variations using a combination of all these methods, the success rate of such followup programmes currently runs at about one planet per ten or more stars surveyed \citep{pont2005a}. To make the most efficient use of followup facilities, more efficient candidate-selection methods are clearly needed.
Our goal is to perform efficient selection using the discovery observations in conjunction with data mining of publicly-available databases of photometry and proper motions, without  having to perform further time-consuming observations. 

Here we present a Bayesian approach to identifying targets whose light curve morphologies and infrared colours are consistent with expectations for main-sequence stars with transiting Jupiter-sized companions. In Sections~\ref{sec:obser} and \ref{sec:photests} we describe the SuperWASP photometry and followup spectroscopy. 
The spectra, augmented by photometric tests for eliminating blends with nearby eclipsing binaries and grazing stellar binaries with observable secondary eclipses,yield spectroscopic and photometric classifications of 67 high-priority transit candidates selected from the 2004 SuperWASP transit survey data. 

In Section~\ref{sec:mcmc} we analyse the light curves of all 67 stars to derive posterior probability distributions for their orbital parameters and radii via Markov-chain Monte-Carlo analysis. 
In Section~\ref{sec:compare} we show that the stellar and planetary parameters obtained for WASP-1 and WASP-2 with this method compare well with results obtained from high-precision transit photometry with larger instruments. In Section~\ref{sec:criteria} we develop candidate selection criteria based on the posterior probability that the companion has a radius appropriate to a planet, and apply them to the SuperWASP sample. An ancillary method of dwarf-giant separation based on the reduced proper motion and the $J-H$ colour is given in Section~\ref{sec:rpm}. 


\begin{table*}
\caption[]{Classification criteria for 67 SuperWASP transit candidates. The codes in the Tel/Inst column are E = OHP 1.93-m with ELODIE, S=OHP 1.93-m with SOPHIE, I = Isaac Newton Telescope with Intermediate-Dispersion Spectrograph, T=Tautenburg 2-m Telescope, W=SuperWASP, X = XO project. Spectroscopic classifications denote whether the spectrum is rotationally broadened (Gaussian $\sigma>8$ km s$^{-1}$), single or double, upper limits on radial-velocity variability, and BisVar= Line-bisector variability. The presence of interstellar sodium or Class III pressure broadening in the wings of strong lines indicates a giant. Photometric classifications are AB=Aperture blend, SE2= Secondary eclipse seen when folded on twice best-fit period, EV2=Ellipsoidal variation seen when folded on twice best-fit period. The values of the main-sequence (MS) prior and the probability that the companion radius is less than $1.5\rmsub{R}{Jup}$ yield the selection criteria in the penultimate column; the final selection including photometric rejection criteria is given in the final column. The acronyms RPM and MCMC denote reduced proper motion and Markov-chain Monte Carlo respectively.}
\label{tab:classify}
\begin{tabular}{lllllrrrr}
\hline\\
1SWASP	&	Tel/Inst	&	Spec.	&	Phot.	&	RPM	&	MS Prior	&	Probability	&	Select?	&	Select?	\\
	&		&	Class	&	Class	&	Class	&	$S=-2\ln$ 	&	($R_p < 1.5 R_{Jup}$)	&	(MCMC 	&	(MCMC 	\\
	&		&		&		&		&	$\mathcal{P}(M_*,R_*)$	&		&	only)	&	\&\ phot)	\\
\hline\\
{\bf Planets:}	&		&		&		&		&		&		&		&		\\
J002040.07+315923.7	&	S	&	Planet (WASP-1)	&	---	&	Dwarf	&	0.98	&	0.989	&	TRUE	&	TRUE	\\
J160211.83+281010.4	&	X	&	Planet (XO-1)	&	---	&	Dwarf	&	0.04	&	0.958	&	TRUE	&	TRUE	\\
J203054.12+062546.4	&	S	&	Planet (WASP-2)	&	---	&	Dwarf	&	0.84	&	1.000	&	TRUE	&	TRUE	\\
\hline\\																	
{\bf Sp. Binaries:}	&		&		&	---	&		&		&		&		&		\\
J130322.00+350525.4	&	I	&	SB1	&	---	&	Dwarf	&	3.87	&	0.445	&	TRUE	&	TRUE	\\
J133339.36+494321.6	&	E	&	SB1	&	---	&	Dwarf	&	0.24	&	0.992	&	TRUE	&	TRUE	\\
J134815.27+464811.0	&	E	&	Broad	&	---	&	Dwarf	&	0.01	&	0.864	&	TRUE	&	TRUE	\\
J174118.30+383656.3	&	S	&	SB1	&	---	&	Dwarf	&	0.86	&	0.983	&	TRUE	&	TRUE	\\
J175401.58+322112.6	&	E	&	SB1 	&	---	&	Dwarf	&	0.00	&	0.995	&	TRUE	&	TRUE	\\
J181858.42+103550.1	&	E	&	Broad	&	---	&	Dwarf	&	0.26	&	0.905	&	TRUE	&	TRUE	\\
J184119.02+403008.4	&	T	&	SB1	&	---	&	Dwarf	&	0.86	&	0.970	&	TRUE	&	TRUE	\\
J203704.92+191525.1	&	T	&	SB1	&	---	&	Dwarf	&	0.08	&	0.999	&	TRUE	&	TRUE	\\
J210009.75+193107.1	&	E	&	Broad	&	---	&	Dwarf	&	0.30	&	0.998	&	TRUE	&	TRUE	\\
J211608.42+163220.3	&	S	&	SB2	&	---	&	Dwarf	&	1.42	&	0.396	&	TRUE	&	TRUE	\\
J212855.03+075753.5	&	I	&	SB1	&	---	&	Dwarf	&	0.11	&	0.312	&	TRUE	&	TRUE	\\
J223320.44+370139.1	&	S	&	SB1	&	---	&	Dwarf	&	0.22	&	0.998	&	TRUE	&	TRUE	\\
J234318.41+295556.5	&	S	&	SB1	&	---	&	Dwarf	&	0.10	&	0.988	&	TRUE	&	TRUE	\\
J010151.11+314254.7	&	S,T	&	SB1 broad	&	---	&	Dwarf	&	17.56	&	0.970	&	FALSE	&	FALSE	\\
J023445.65+251244.0	&	T	&	SB1	&	---	&	Dwarf	&	7.37	&	0.034	&	FALSE	&	FALSE	\\
J031103.19+211141.4	&	S	&	Broad	&	---	&	Dwarf	&	10.52	&	0.000	&	FALSE	&	FALSE	\\
J051221.34+300634.9	&	S	&	Multiple	&	---	&	Dwarf	&	7.36	&	0.001	&	FALSE	&	FALSE	\\
J115718.66+261906.1	&	I	&	SB1	&	---	&	Dwarf	&	0.42	&	0.074	&	FALSE	&	FALSE	\\
J141558.71+400026.7	&	E	&	Broad	&	---	&	Dwarf	&	0.12	&	0.002	&	FALSE	&	FALSE	\\
J152131.01+213521.3	&	I	&	SB1	&	---	&	Dwarf	&	2.21	&	0.000	&	FALSE	&	FALSE	\\
J153135.51+305957.1	&	E	&	Broad	&	---	&	Dwarf	&	11.67	&	0.000	&	FALSE	&	FALSE	\\
J165949.13+265346.1	&	S	&	Broad	&	---	&	Dwarf	&	8.41	&	0.032	&	FALSE	&	FALSE	\\
J172549.13+502206.4	&	E	&	Broad	&	---	&	Dwarf	&	2.66	&	0.150	&	FALSE	&	FALSE	\\
J173650.17+105557.9	&	E	&	Broad	&	---	&	Dwarf	&	0.60	&	0.000	&	FALSE	&	FALSE	\\
J173748.98+471348.7	&	I	&	SB1	&	---	&	Dwarf	&	7.84	&	1.000	&	FALSE	&	FALSE	\\
J174327.81+582512.7	&	E	&	Broad	&	---	&	Dwarf	&	14.02	&	0.002	&	FALSE	&	FALSE	\\
J175511.09+134731.5	&	E	&	Broad	&	---	&	Dwarf	&	10.39	&	0.042	&	FALSE	&	FALSE	\\
J175620.84+253625.7	&	T	&	SB2	&	---	&	Dwarf	&	3.35	&	0.005	&	FALSE	&	FALSE	\\
J180010.55+214510.2	&	T	&	SB1	&	---	&	Giant	&	0.13	&	0.010	&	FALSE	&	FALSE	\\
J182620.36+475902.8	&	E	&	SB2	&	---	&	Giant	&	90.38	&	0.000	&	FALSE	&	FALSE	\\
J202824.02+192310.2	&	E	&	Broad	&	---	&	Dwarf	&	0.86	&	0.000	&	FALSE	&	FALSE	\\
J203906.39+171345.9	&	E	&	SB1	&	---	&	Dwarf	&	4.67	&	0.015	&	FALSE	&	FALSE	\\
J215802.14+253006.1	&	S	&	SB2	&	---	&	Dwarf	&	32.53	&	0.906	&	FALSE	&	FALSE	\\
J231533.56+232637.5	&	T	&	SB1	&	---	&	Dwarf	&	59.89	&	0.017	&	FALSE	&	FALSE	\\
\hline\\																
\end{tabular}
\end{table*}

\begin{table*}
\contcaption{}	
\begin{tabular}{lllllrrrr}
\hline\\
1SWASP	&	Tel/Inst	&	Spec.	&	Phot.	&	RPM	&	MS Prior	&	Probability	&	Select?	&	Select?	\\
	&		&	Class	&	Class	&	Class	&	$S=-2\ln$ 	&	($R_p < 1.5 R_{Jup}$)	&	(MCMC 	&	(MCMC 	\\
	&		&		&		&		&	$\mathcal{P}(M_*,R_*)$	&		&	only)	&	\&\ phot)	\\
\hline\\
{\bf Phot. Binaries:}	&		&		&		&		&		&		&		&		\\
J003039.21+205719.1	&	T,W	&	SB2	&	SE2	&	Dwarf	&	1.60	&	0.022	&	FALSE	&	FALSE	\\
J015711.29+303447.7	&	T	&	SB2	&	SE2	&	Dwarf	&	1.42	&	0.927	&	TRUE	&	FALSE	\\
J160242.43+290850.1	&	W	&	---	&	SE2	&	Giant	&	16.32	&	0.000	&	FALSE	&	FALSE	\\
J165423.72+241335.7	&	W	&	---	&	SE2	&	Giant	&	2.88	&	0.847	&	TRUE	&	FALSE	\\
J183104.01+323942.7	&	S	&	SB2	&	SE2	&	Dwarf	&	0.68	&	0.997	&	TRUE	&	FALSE	\\
J183805.57+423432.3	&	W	&	---	&	SE2	&	Giant	&	4.08	&	0.995	&	TRUE	&	FALSE	\\
J205308.03+192152.7	&	S	&	SB2	&	SE2	&	Dwarf	&	3.22	&	1.000	&	TRUE	&	FALSE	\\
J210909.05+184950.9	&	W	&	---	&	EV2	&	Dwarf	&	0.36	&	0.991	&	TRUE	&	FALSE	\\
J222353.83+412813.5	&	W	&	---	&	EV2	&	Dwarf	&	53.32	&	0.000	&	FALSE	&	FALSE	\\
J223651.20+221000.8	&	S	&	SB2	&	SE2	&	Giant	&	1.13	&	0.021	&	FALSE	&	FALSE	\\
\hline\\
{\bf Aperture blends:}	&		&		&		&		&		&		&		&		\\
J025500.31+281134.0	&	S,T	&	$<10$ m s$^{-1}$	&	AB	&	Giant	&	15.19	&	0.000	&	FALSE	&	FALSE	\\
J133156.81+460026.6	&	W	&	---	&	AB	&	Giant	&	82.69	&	0.000	&	FALSE	&	FALSE	\\
J161644.68+200806.8	&	W	&	---	&	AB	&	Giant	&	79.66	&	0.000	&	FALSE	&	FALSE	\\
J181454.99+391146.0	&	W	&	---	&	AB	&	Giant	&	2.55	&	1.000	&	TRUE	&	FALSE	\\
J184303.62+462656.4	&	S	&	I/S NaI, ClassIII	&	AB	&	Giant	&	60.26	&	0.971	&	FALSE	&	FALSE	\\
J204142.49+075051.5	&	W	&	---	&	AB	&	Giant	&	1.54	&	1.000	&	TRUE	&	FALSE	\\
J204712.42+202544.5	&	W,T	&	$<1$ km s$^{-1}$	&	AB	&	Giant	&	1.93	&	1.000	&	TRUE	&	FALSE	\\
J204745.08+103347.9	&	W	&	---	&	AB	&	Giant	&	65.54	&	0.780	&	FALSE	&	FALSE	\\
J213416.37+205644.4	&	W,T	&	---	&	AB	&	Dwarf	&	35.47	&	0.000	&	FALSE	&	FALSE	\\
J215226.17+331424.7	&	W	&	---	&	AB	&	Dwarf	&	52.29	&	0.036	&	FALSE	&	FALSE	\\
J222317.60+130125.8	&	S	&	I/S NaI, ClassIII	&	AB	&	Giant	&	38.09	&	0.994	&	FALSE	&	FALSE	\\
J223809.90+401038.1	&	W	&	---	&	AB	&	Giant	&	5.55	&	1.000	&	FALSE	&	FALSE	\\
J224104.57+363648.3	&	W	&	---	&	AB	&	Giant	&	43.77	&	0.994	&	FALSE	&	FALSE	\\
\hline\\																	
{\bf Other blends, giants:}	&		&		&		&		&		&		&		&		\\
J161732.90+242119.0	&	S,E	&	$<10$ m s$^{-1}$	&	---	&	Dwarf	&	0.53	&	1.000	&	TRUE	&	TRUE	\\
J210318.01+080117.8	&	S	&	I/S NaI, BisVar	&	---	&	Dwarf	&	0.26	&	0.983	&	TRUE	&	TRUE	\\
J005225.90+203451.2	&	S,T	&	$<10$ m s$^{-1}$	&	---	&	Dwarf	&	14.03	&	0.008	&	FALSE	&	FALSE	\\
J181252.03+461851.6	&	S,E	&	$<10$ m s$^{-1}$	&	---	&	Dwarf	&	2.46	&	0.036	&	FALSE	&	FALSE	\\
J204125.28+163911.8	&	S	&	ClassIII	&	---	&	Giant	&	37.85	&	1.000	&	FALSE	&	FALSE	\\
J205027.33+064022.9	&	S	&	$<10$ m s$^{-1}$	&	---	&	Dwarf	&	33.23	&	0.986	&	FALSE	&	FALSE	\\
J214151.03+260158.5	&	S	&	---	&	---	&	Giant	&	26.81	&	1.000	&	FALSE	&	FALSE	\\
\hline\\																	
\end{tabular}
\end{table*}	

\section{Observations}
\label{sec:obser}

Among the 109 SuperWASP transit candidates selected for potential spectroscopic followup,  67 stars were examined in sufficient detail to establish their nature. These 67 objects are listed
in Table~\ref{tab:classify} together with their spectroscopic and photometric classifications. They were initially selected for followup using criteria described by \citet{cameron2006}, \citet{christian2006}, 
\citet{street2007} and \citet{lister2007}. Among them are the three known planet host stars XO-1 \citep{mccullough2006}, WASP-1 and WASP-2 \citep{cameron2007}.  The remaining 64 were eliminated as planet candidates using the spectroscopic and photometric criteria described here and in Section~\ref{sec:photests}.

\subsection{SuperWASP photometry}

The light curves analysed in this paper were secured between 2004 May and September using the SuperWASP wide-field camera array. The SuperWASP instrument and data pipeline are described in full by \citet{pollacco2006}. During 2004, the instrument comprised five f/1.8 Canon lenses of 200-mm focal length backed by Andor CCD arrays of $2048\times 2048$ 13 $\mu$m pixels. Each camera imaged an area of sky 7.8$^\circ$ square. The automated observing routine followed a raster pattern sweeping from 3.5 hours east to 3.5 hours west of the meridian, returning to a given field every 6 minutes on average. The light curves of most targets comprise some 3000 observations secured over a period of 100 to 150 days. 

The ensemble photometry from each camera on each night was calibrated and corrected for extinction, colour and zero-point terms using networks of local secondary standards. Additional patterns of systematic error affecting all the stars were identified and removed using the SysRem algorithm of \citet{tamuz2005}. The search for transit-like events and selection of astrophysically-plausible transit candidates was performed using the methodology of \citet{cameron2006}, backed up by visual inspection of the light curves to verify the reality of the transits as described by \citet{christian2006}.

In total these procedures yielded a ``long-list" of 109 transit candidates with periods less than 5 days, from which obvious impostors such as ellipsoidal variables and objects showing clear secondary eclipses on the best-fitting period had been eliminated.

\subsection{OHP 1.93-m spectroscopy}

The initial spectroscopic followup of transit candidates identified from the 2004 SuperWASP survey was carried out during three 4-night observing runs on the 1.93-m telescope at l'Observatoire de Haute-Provence (OHP). The first two runs, in 2006 May and June, utilised the ELODIE spectrograph and radial-velocity software \citep{bouchy2006sophie}. ELODIE was replaced by its successor SOPHIE in time for the third run of the series, in 2006 August. Both instruments were used in their ``high-efficiency" modes, which cover the wavelength range from 387 to 694 nm with resolving power $\lambda/\Delta\lambda=35000$. Radial velocities are computed at the telescope, using cross-correlation with a mask spectrum, using ThAr arc spectra to monitor the stability of the instrument. The typical radial-velocity precisions were of order 40 m s$^{-1}$ for narrow-lined objects observed with ELODIE, and 12 m s$^{-1}$ with SOPHIE. The cross-correlation functions allow several types of astrophysical false positive to be identified from a single exposure. Targets were rejected if the first-epoch spectrum revealed them to be double-lined or rapidly rotating (and hence presumably tidally locked by a stellar-mass companion). Examples of such cross-correlation functions are illustrated in Fig. 1 of \citet{pont2005a}. Velocity shifts of more than a few hundred m s$^{-1}$ between the first spectrum and a second measurement on a subsequent night also led to rejection as a stellar binary.

As Table~\ref{tab:classify} shows, 14 targets were eliminated as broad, single-lined or double-lined binaries during the ELODIE runs.  12 stars were identified as binaries with SOPHIE, while a further 10
were found to be either giants (deduced visually from absence of pressure-broadening wings on the Na I D and Mg I b lines, and/or from the presence of narrow foreground  interstellar sodium absorption), or to exhibit no detectable radial velocity variation (suggesting blending with a nearby eclipsing binary).

\subsection{Tautenburg spectroscopy}

Further spectra of several targets were secured using the coud\'{e} echelle spectrograph on the 2-m Alfred-Jensch-Teleskop at the Th{\"u}ringer Landessternwarte Tautenburg (TLS) over the winter of 2006-7. A 2-arcsec slit width was used, yielding a spectral resolving power $\lambda/\Delta\lambda=35000$. For most observations, velocities were obtained by cross-correlation referenced to the ThAr wavelength calibration, using telluric lines as a secondary reference to monitor and correct possible instrumental shifts. The overall velocity precision was typically of order 200 m s$^{-1}$. The rejection criteria were similar to those described above for the OHP observations. Six additional stars were eliminated as binaries
on the basis of these observations.

\subsection{INT spectroscopy}

Spectra were secured of several previously-unobserved targets using the Intermediate-Dispersion Spectrograph (IDS) on the 2.5-m Isaac Newton Telescope (INT) at the Observatorio del Roque de los Muchachos, La Palma, during 2007 April. Similar rejection criteria were applied. Five stars were found to be single-lined binaries.

\section{Further photometric tests}
\label{sec:photests}

\begin{figure}
\begin{center}
\psfig{figure=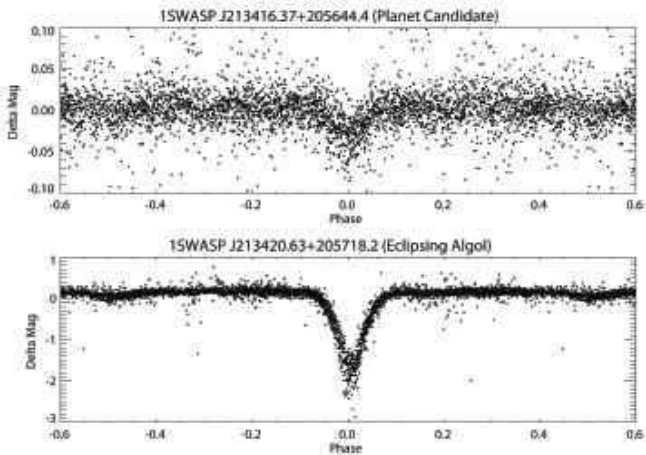,width=8.5cm}
\caption[]{Blended eclipse of the planet candidate 1SWASP J213416.37+205644.4. The top panel shows the low amplitude eclipse induced in the lightcurve by the eclipsing Algol BQ Peg at a distance of 69 arcsec. The bottom panel shows the lightcurve for BQ Peg itself. The lack of a sharp minimum on the primary eclipse is due to the poorer data quality as the flux approaches the SuperWASP faint limit at the $V=16.7$ primary minimum. Both are phase folded on the BQ Peg period of 1.574520 days.}
\label{fig:planet_blend}
\end{center}
\end{figure}

\begin{figure}
\begin{center}
\psfig{figure=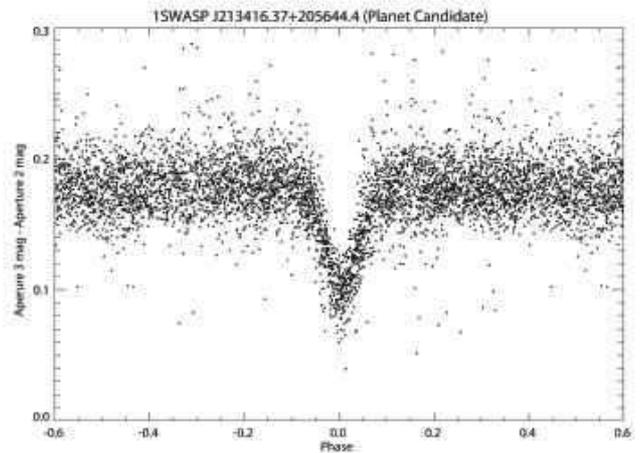,width=8.5cm}
\caption[]{Aperture-blend test for planet candidate 1SWASP J213416.37+205644.4. The plot shows the magnitude in the main photometric aperture minus that in the larger-outer aperture, phase folded on the determined period of the planet candidate. A V-shaped eclipse is clearly seen indicating this object is blended with a resolved eclipsing binary system, in this case BQ Peg.}
\label{fig:ap_blend}
\end{center}
\end{figure}

\subsection{Resolved aperture blends}
\label{sec:apblend}

Eclipsing binary systems blended with nearby bright, non-variable stars represent the most common type of mimic in planetary transit surveys \citep{pont2005a}. The WASP data reduction pipeline performs photometry in three concentric circular apertures centred on each object, with radii of 2.5, 3.5 and 4.5 13.5-arcsec pixels. In cases where an eclipsing binary system lies just outside the main (3.5-pixel radius) photometric aperture of a non-variable star, the two systems will contribute different amounts of light to different apertures. Light leaking from the wings of the binary's image into the apertures of the non-variable star may mimic a shallow transit at times when the binary is in eclipse. The amount of leakage 
is greatest in the outer apertures.  As a consequence, the depth of the ``transit'' in the non-variable object is greatest when measured in the outer apertures.

The eclipsing binary BQ Pegasi is a typical example. This is an Algol system with primary eclipses 2.6 mag deep, located 69 arcsec away from a 12th magnitude non-variable star (1SWASP J213416.37+205644.4). The centre-to-centre separation of these two systems is approximately 5 pixels causing a small amount of overlap between the photometric apertures. This overlap was sufficient to induce a ~2\% dip in brightness in the non-variable star, closely mimicking the photometric signal of a transiting planet with a period of 1.57455 days -- identical to BQ Peg. As a result the non-varying star was flagged by the transit detection algorithm as a high-priority planet candidate. The planet candidate lightcurve is shown in Figure~\ref{fig:planet_blend} together with the lightcurve for BQ Peg. We see that the phase of the false transit, folded on the period of BQ Peg, matches that of the binary's primary eclipse. Without knowledge of the variable nature of the blending system this candidate would be difficult to eliminate without further followup. However, a plot of the difference in magnitudes between the main and outer photometric apertures (Figure~\ref{fig:ap_blend}), phase folded on the orbital period determined for the planet candidate, shows a clear V-shaped eclipse indicating that the star is blended. A non-blended system would be not be expected to produce a change in the flux ratio between different apertures in this situation. 

Thirteen objects failed the photometric aperture-blend test. Five among these were also observed spectroscopically. Two were found to be giants; of the two that were observed more than once, neither showed significant radial-velocity variability.

\subsection{Ellipsoidal variables and secondary eclipses}
\label{sec:sececl}

Because the transit search was restricted to periods less than 5 days,
a few objects were detected in which secondary eclipses and/or ellipsoidal variability
were present, but in which the best-fitting period found by the transit-search algorithm was half the true period. To guard against this possibility we examined the light curves of all candidates phase-folded on twice the best-fitting period.

Two were eliminated photometrically as exhibiting ellipsoidal variability on twice the orbital period found by the transit detection software. A further seven objects were found to exhibit unequal depths or displacements in phase between odd-numbered and even-numbered transits, indicating that the periods were indeed twice those found by the transit-search software, and that secondary eclipses were present. The five among these that were also observed spectroscopically were all found to be double-lined binaries.

\section{Markov-Chain Monte-Carlo modelling}
\label{sec:mcmc}

Markov-chain Monte-Carlo (MCMC) methods are rapidly gaining popularity as an efficient and informative means of solving of multivariate parameter-fitting problems in astronomy and many other branches of science. They provide not just a means 
of optimising the fit of a model to data, but their mode of operation allows a precise exploration of the joint posterior probability distribution of the fitted parameters. \citet{tegmark2004} give a clear description of the technique and its use for deriving limits on cosmological parameters from the spatial power spectrum of the cosmic microwave background. \citet{ford2006} and \citet{gregory2007} have applied MCMC to the derivation of orbital parameters and detection  of additional planets from radial-velocity curves. \citet{holman2006} have shown Markov-Chain Monte-Carlo (MCMC) methods also to be particularly well-suited to the problem of deriving the physical parameters of star-planet systems by optimising model fits to the light curves of transiting exoplanets. 

\begin{figure*}
\begin{center}
\psfig{figure=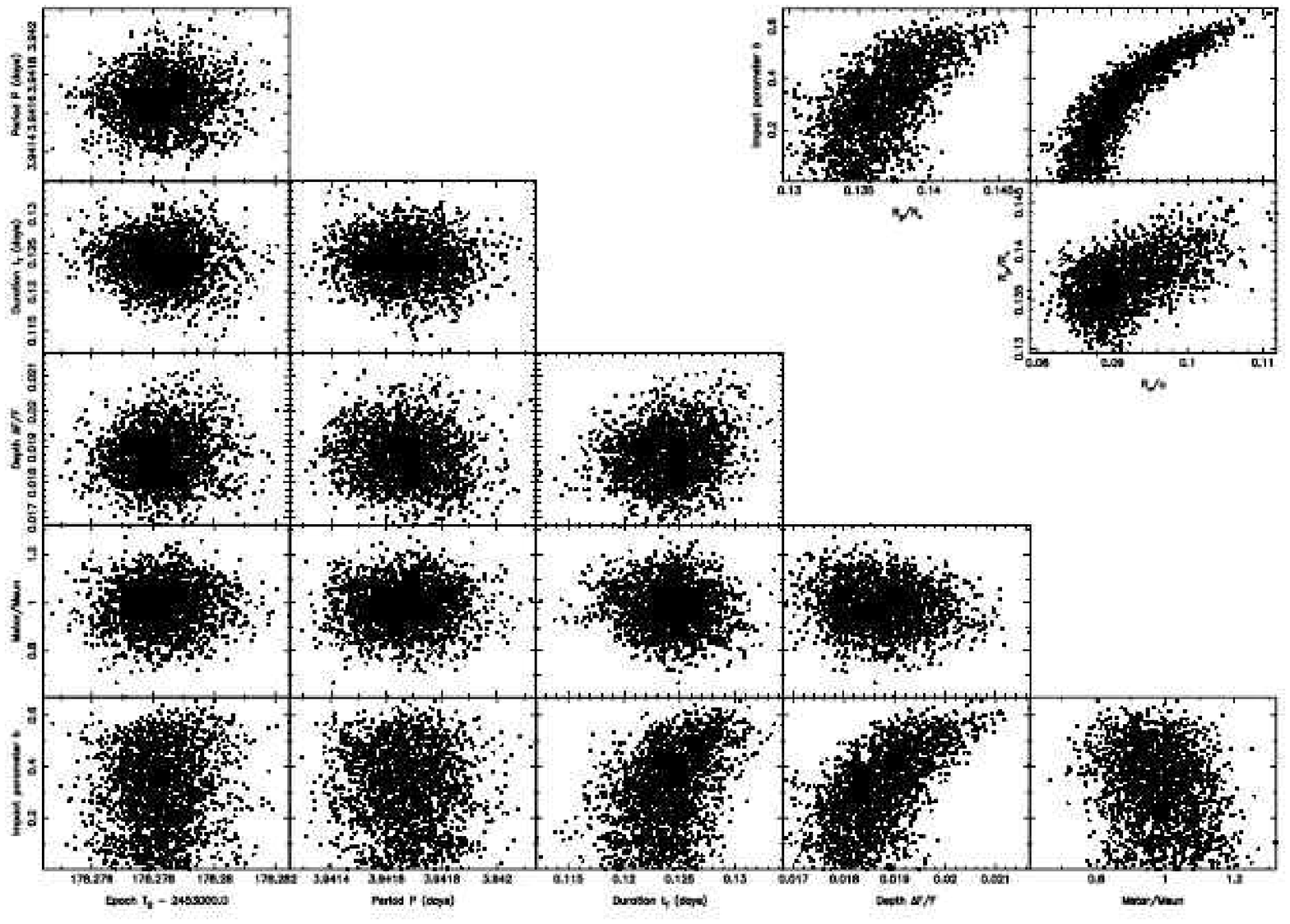,width=17cm}
\caption[]{The lower triangular matrix shows the correlation diagram for the six proposal parameters  $\{T_0,P,\Delta F,t_T,b,M_*\}$ for an MCMC analysis of the 2004 SuperWASP light curve of the planet host star XO-1. The MCMC run imposed Bayesian priors on both the stellar mass and the stellar radius, derived from the $J-H$ colour and the main-sequence mass-radius relation. The shape of the transit profile allows a wide range of intermediate values for the impact parameter $b$. Nonetheless, the majority of the proposal parameters are very close to being  mutually uncorrelated, with the exception of $b$ and $\Delta F$, which are correlated through the effect of limb darkening. The upper triangular matrix shows the mutual correlations among the physical parameters $b$ and $R_*/a$ and $R_p/R_*$. The well-known correlation between $b$ and $R_*/a$ is very much stronger than those seen among the six observational parameters that we have adopted here.}
\label{fig:plotmatrix}
\end{center}
\end{figure*}

In our implementation of MCMC, we assume the planet's orbit to be circular. We characterise the system using the six parameters $\{T_0,P,\Delta F,t_T,b,M_*\}$. Here $T_0$ is the epoch of mid-transit, $P$ is the orbital period, $\Delta F$ is the fractional flux deficit that would be observed during transit in the absence of limb darkening, $t_T$ is the total duration of the transit from first to last contact, $b$ is the impact parameter of the planet's path across the stellar disc in units of the primary's radius, and $M_*$ is the mass of the primary in solar units. These six quantities constitute the ``proposal parameters", which are allowed to perform a random walk through parameter space, generating a cloud of points that map out the joint posterior probability distribution (Fig.~\ref{fig:plotmatrix}). The best-fitting values of the proposal parameters are listed for all 67 stars in Table~\ref{tab:bestparms}.

At each step in the MCMC procedure, each proposal parameter is perturbed from its previous value by a small random amount:
\begin{eqnarray*}
T_{0,i} & = & T_{0,i-1}+\sigma_{T_0}G(0,1)f\\
P_{i} & = & P_{i-1}+\sigma_{P}G(0,1)f\\
\Delta F_{i} & = & \Delta F_{i-1}+\sigma_{\Delta F}G(0,1)f\\
t_{T,i} & = & t_{T,i-1}+\sigma_{t_T}G(0,1)f\\
b_{i} & = & T_{0,i-1}+\sigma_{b}G(0,1)f\\
M_{*,i} & = & M_{*,i-1}+\sigma_{M}G(0,1)f
\end{eqnarray*}
where $G(0,1)$ is a random Gaussian deviate with mean zero and unit standard deviation. The scale factor $f$ is an adaptive step-size controller whose value is initially set to 0.5, but which evolves together with the estimated parameter uncertainties as described in Section~\ref{sec:msprior} below.

The first four parameters ($T_0$, $P$, $\Delta F$ and $t_T$) are more or less directly measurable from the folded light curve. Their initial values and their associated one-sigma uncertainties are taken directly from the results of the accelerated Box Least-Squares algorithm of \citet{cameron2006}. The impact parameter is given an initial value $b_0=0.5$ and a one-sigma uncertainty $\sigma_b = 0.05$. The stellar mass $M_*$ is initially set to the value $M_{0}$ derived from the $J-H$ colour using the calibration described in Appendix~\ref{sec:jhcalib}, and assigned an arbitrary but plausible one-sigma uncertainty $\sigma_M=0.1M_{0}$. 

Once the physical parameters $R_*$, $R_p$, $a$ and $\cos i$ have been derived from the proposal parameters $\Delta F$, $t_T$, $b$ and $P$  using the relationships presented in Appendix~\ref{sec:transform}, it is straightforward to compute the projected separation of centres (in units of the primary radius) at any time $t_j$ of observation
\begin{equation}
z(t_j)=\frac{\sin^2\phi_j+(bR_*/a)^2\cos^2\phi_j}{R_*/a}.
\end{equation}
The orbital phase angle at time $t_j$ is $\phi_j =2\pi( t_j-T_0)/P$. 

\begin{table*}
\caption[]{Observable and physical parameters for the 67 stars in the
sample, computed by optimising the posterior probability $Q$ incorporating
the main-sequence prior constraint on the primary mass and radius.}
\label{tab:bestparms}
\begin{tabular}{lrrrrrrrr}
\hline\\
1SWASP	&	Transit	&	Orbital	&	Transit	&	Transit	&	Impact	&	Stellar	&	Primary	&	Secondary	\\
	&	Epoch $T_0$	&	period $P$	&	duration	&	depth	&	parameter	&	Mass	&	radius	&	radius	\\
	&	(HJD)	&	(days)	&	$t_T/P$	&	$\Delta F$	&	$b$	&	$M_*/M_\odot$	&	$R_*/R_\odot$	&	$R_2/\rmsub{R}{Jup}$	\\
\hline\\
{\bf Planets:}	&		&		&		&		&		&		&		&		\\
J002040.07+315923.7	&	3219.5240	&	2.52065	&	0.055	&	0.010	&	0.028	&	1.24	&	1.38	&	1.36	\\
J160211.83+281010.4	&	3178.2780	&	3.94164	&	0.032	&	0.019	&	0.402	&	0.99	&	0.97	&	1.30	\\
J203054.12+062546.4	&	3216.7138	&	2.15202	&	0.034	&	0.015	&	0.696	&	0.84	&	0.83	&	1.02	\\
\hline\\
{\bf Sp. Binaries:}	&		&		&		&		&		&		&		&		\\
J130322.00+350525.4	&	3167.4755	&	2.67382	&	0.055	&	0.014	&	0.038	&	1.05	&	1.28	&	1.48	\\
J133339.36+494321.6	&	3175.1566	&	3.64942	&	0.038	&	0.009	&	0.021	&	1.33	&	1.19	&	1.08	\\
J134815.27+464811.0	&	3169.5603	&	1.05603	&	0.056	&	0.009	&	0.807	&	1.04	&	1.04	&	0.98	\\
J174118.30+383656.3	&	3197.8605	&	4.24481	&	0.037	&	0.010	&	0.280	&	1.28	&	1.32	&	1.28	\\
J175401.58+322112.6	&	3194.0863	&	1.94918	&	0.050	&	0.012	&	0.537	&	1.13	&	1.10	&	1.15	\\
J181858.42+103550.1	&	3170.8197	&	2.46530	&	0.056	&	0.010	&	0.113	&	1.41	&	1.39	&	1.34	\\
J184119.02+403008.4	&	3188.2544	&	3.73765	&	0.040	&	0.013	&	0.058	&	1.06	&	1.15	&	1.29	\\
J203704.92+191525.1	&	3215.3230	&	1.68008	&	0.045	&	0.009	&	0.762	&	1.12	&	1.11	&	1.02	\\
J210009.75+193107.1	&	3194.0029	&	3.05562	&	0.041	&	0.007	&	0.516	&	1.61	&	1.40	&	1.13	\\
J211608.42+163220.3	&	3219.5782	&	3.46829	&	0.031	&	0.015	&	0.842	&	1.23	&	1.35	&	1.60	\\
J212855.03+075753.5	&	3220.9833	&	4.68994	&	0.023	&	0.030	&	0.757	&	0.95	&	0.92	&	1.56	\\
J223320.44+370139.1	&	3240.4446	&	1.87478	&	0.050	&	0.010	&	0.565	&	1.29	&	1.17	&	1.11	\\
J234318.41+295556.5	&	3245.1886	&	4.24098	&	0.030	&	0.021	&	0.008	&	0.78	&	0.85	&	1.20	\\
J010151.11+314254.7	&	3230.3097	&	3.65098	&	0.050	&	0.014	&	0.010	&	0.76	&	1.28	&	1.45	\\
J023445.65+251244.0	&	3234.8368	&	1.55373	&	0.082	&	0.018	&	0.020	&	0.93	&	1.26	&	1.65	\\
J031103.19+211141.4	&	3250.5698	&	2.72925	&	0.063	&	0.031	&	0.020	&	0.94	&	1.36	&	2.35	\\
J051221.34+300634.9	&	3253.3499	&	1.23758	&	0.085	&	0.028	&	0.657	&	1.03	&	1.36	&	2.22	\\
J115718.66+261906.1	&	3156.3795	&	2.45397	&	0.058	&	0.014	&	0.066	&	1.36	&	1.38	&	1.60	\\
J141558.71+400026.7	&	3170.0586	&	1.64277	&	0.055	&	0.050	&	0.995	&	1.33	&	1.29	&	3.95	\\
J152131.01+213521.3	&	3154.5622	&	4.01493	&	0.043	&	0.026	&	0.082	&	1.19	&	1.32	&	2.08	\\
J153135.51+305957.1	&	3181.9327	&	4.46754	&	0.047	&	0.029	&	0.070	&	0.94	&	1.40	&	2.31	\\
J165949.13+265346.1	&	3200.5994	&	2.68305	&	0.042	&	0.023	&	0.885	&	1.11	&	1.48	&	2.35	\\
J172549.13+502206.4	&	3198.8106	&	2.27129	&	0.044	&	0.025	&	0.785	&	1.01	&	1.19	&	1.84	\\
J173650.17+105557.9	&	3193.1694	&	3.44102	&	0.041	&	0.030	&	0.037	&	0.94	&	1.03	&	1.74	\\
J173748.98+471348.7	&	3182.5662	&	3.33791	&	0.046	&	0.008	&	0.025	&	0.86	&	1.19	&	1.05	\\
J174327.81+582512.7	&	3213.7625	&	2.84493	&	0.064	&	0.013	&	0.016	&	1.01	&	1.53	&	1.69	\\
J175511.09+134731.5	&	3203.1171	&	2.44367	&	0.057	&	0.019	&	0.602	&	1.03	&	1.43	&	1.92	\\
J175620.84+253625.7	&	3204.0353	&	4.41602	&	0.034	&	0.031	&	0.667	&	1.05	&	1.28	&	2.18	\\
J180010.55+214510.2	&	3198.2836	&	4.41513	&	0.023	&	0.051	&	0.751	&	0.67	&	0.75	&	1.66	\\
J182620.36+475902.8	&	3205.5267	&	3.04416	&	0.086	&	0.068	&	0.935	&	0.05	&	0.91	&	2.99	\\
J202824.02+192310.2	&	3211.1395	&	1.25788	&	0.092	&	0.018	&	0.016	&	1.30	&	1.36	&	1.76	\\
J203906.39+171345.9	&	3209.4011	&	2.69705	&	0.045	&	0.021	&	0.758	&	1.16	&	1.42	&	1.99	\\
J215802.14+253006.1	&	3215.6417	&	1.51272	&	0.046	&	0.018	&	0.332	&	0.75	&	0.67	&	0.87	\\
J231533.56+232637.5	&	3225.8489	&	4.56289	&	0.061	&	0.016	&	0.003	&	0.35	&	1.39	&	1.74	\\
\hline\\
\end{tabular}															\end{table*}															

\begin{table*}
\contcaption{}
\label{}
\begin{tabular}{lrrrrrrrr}
\hline\\
1SWASP	&	Transit	&	Orbital	&	Transit	&	Transit	&	Impact	&	Stellar	&	Primary	&	Secondary	\\
	&	Epoch $T_0$	&	period $P$	&	duration	&	depth	&	parameter	&	Mass	&	radius	&	radius	\\
	&	(HJD)	&	(days)	&	$t_T/P$	&	$\Delta F$	&	$b$	&	$M_*/M_\odot$	&	$R_*/R_\odot$	&	$R_2/\rmsub{R}{Jup}$	\\
\hline\\
{\bf Phot. Binaries:}	&		&		&		&		&		&		&		&		\\
J003039.21+205719.1	&	3230.6250	&	4.56617	&	0.028	&	0.020	&	0.792	&	1.21	&	1.34	&	1.86	\\
J015711.29+303447.7	&	3237.7106	&	2.04321	&	0.056	&	0.012	&	0.428	&	1.05	&	1.17	&	1.27	\\
J160242.43+290850.1	&	3187.2622	&	1.30433	&	0.082	&	0.044	&	0.681	&	0.66	&	1.13	&	2.30	\\
J165423.72+241335.7	&	3188.8269	&	2.57122	&	0.045	&	0.031	&	0.030	&	0.60	&	0.80	&	1.36	\\
J183104.01+323942.7	&	3210.9015	&	2.37843	&	0.040	&	0.008	&	0.769	&	1.30	&	1.33	&	1.18	\\
J183805.57+423432.3	&	3199.0823	&	3.51709	&	0.036	&	0.016	&	0.046	&	0.59	&	0.82	&	0.99	\\
J205308.03+192152.7	&	3204.4935	&	1.67630	&	0.076	&	0.005	&	0.008	&	1.22	&	1.41	&	1.00	\\
J210909.05+184950.9	&	3216.9989	&	2.91766	&	0.044	&	0.007	&	0.470	&	1.70	&	1.45	&	1.16	\\
J222353.83+412813.5	&	3239.3999	&	1.55304	&	0.130	&	0.038	&	0.112	&	0.48	&	1.51	&	2.87	\\
J223651.20+221000.8	&	3205.6264	&	3.22659	&	0.032	&	0.056	&	0.865	&	0.77	&	0.91	&	2.42	\\
\hline\\
{\bf Aperture blends:}	&		&		&		&		&		&		&		&		\\
J025500.31+281134.0	&	3237.4163	&	2.16594	&	0.068	&	0.029	&	0.034	&	0.70	&	1.14	&	1.88	\\
J133156.81+460026.6	&	3166.2329	&	3.16678	&	0.091	&	0.063	&	0.072	&	0.08	&	0.91	&	2.22	\\
J161644.68+200806.8	&	3243.4648	&	3.96830	&	0.069	&	0.178	&	0.994	&	0.10	&	0.85	&	4.90	\\
J181454.99+391146.0	&	3192.6921	&	1.10269	&	0.067	&	0.018	&	0.081	&	0.47	&	0.65	&	0.85	\\
J184303.62+462656.4	&	3192.1856	&	3.33871	&	0.060	&	0.020	&	0.066	&	0.22	&	0.94	&	1.30	\\
J204142.49+075051.5	&	3207.5200	&	2.76243	&	0.037	&	0.008	&	0.110	&	0.59	&	0.75	&	0.65	\\
J204712.42+202544.5	&	3208.9201	&	2.61303	&	0.041	&	0.020	&	0.004	&	0.60	&	0.76	&	1.06	\\
J204745.08+103347.9	&	3200.9224	&	3.23610	&	0.062	&	0.021	&	0.061	&	0.18	&	0.88	&	1.23	\\
J213416.37+205644.4	&	3220.5575	&	1.57455	&	0.105	&	0.019	&	0.013	&	0.63	&	1.42	&	1.89	\\
J215226.17+331424.7	&	3226.5888	&	1.06591	&	0.149	&	0.018	&	0.039	&	0.41	&	1.34	&	1.75	\\
J222317.60+130125.8	&	3227.4250	&	1.61385	&	0.085	&	0.016	&	0.060	&	0.39	&	1.00	&	1.23	\\
J223809.90+401038.1	&	3240.3573	&	1.40652	&	0.065	&	0.016	&	0.023	&	0.52	&	0.77	&	0.94	\\
J224104.57+363648.3	&	3245.1212	&	1.73399	&	0.081	&	0.019	&	0.100	&	0.31	&	0.92	&	1.24	\\
\hline\\
{\bf Other blends, giants:}	&		&		&		&		&		&		&		&		\\
J161732.90+242119.0	&	3182.9227	&	1.45378	&	0.055	&	0.012	&	0.436	&	0.72	&	0.82	&	0.87	\\
J210318.01+080117.8	&	3214.3962	&	1.22395	&	0.077	&	0.013	&	0.255	&	1.06	&	1.10	&	1.22	\\
J005225.90+203451.2	&	3230.5589	&	1.71894	&	0.059	&	0.031	&	0.869	&	0.81	&	1.27	&	2.36	\\
J181252.03+461851.6	&	3210.2889	&	2.52501	&	0.039	&	0.033	&	0.831	&	0.91	&	1.10	&	1.98	\\
J204125.28+163911.8	&	3214.3614	&	1.22154	&	0.096	&	0.006	&	0.027	&	0.38	&	0.97	&	0.74	\\
J205027.33+064022.9	&	3209.8236	&	1.22927	&	0.125	&	0.007	&	0.007	&	0.74	&	1.58	&	1.31	\\
J214151.03+260158.5	&	3238.9321	&	1.82588	&	0.064	&	0.007	&	0.031	&	0.38	&	0.84	&	0.69	\\
\hline\\
\end{tabular}															\end{table*}

We compute the flux deficit at all observed orbital phases using the algorithm of \citet{mandel2002} for small planets with the 4-coefficient limb-darkening model of \citet{claret2000}. The four limb-darkening coefficients are interpolated from Claret's tables for the $R$ band (whose mean wavelength approximates the effective wavelength of the unfiltered WASP instrumental bandpass) using the stellar effective temperature determined from the $J-H$ colour via equation~\ref{eq:jh_teff}. We adopted surface gravity $\log g=4.5$, a microturbulent velocity $\rmsub{\upsilon}{turb}=2$ km s$^{-1}$, and a metallicity $[Fe/H]=0.1$ on the grounds that planet-bearing stars tend to have heavy-element abundances slightly above solar.
After converting these flux deficits to model magnitudes $\mu_j$ relative to the flux received from the system outside transit, we compute the zero-point offset from the observed magnitudes $m_j$:
\begin{equation}
\Delta m =\frac{ \sum_j (m_j-\mu_j)w_j}{\sum_j w_j},
\end{equation}
where the observational errors $\sigma_j$ define the inverse-variance weights $w_j=1/\sigma^2_j$. We thus obtain the fitting statistic for the set of model parameters 
pertaining to the $i$th step of the Markov chain:
\begin{equation}
\chi^2_i(T_0,P,\Delta F,t_T,b,M_*)=\sum_j\frac{(m_j-\mu_j-\Delta m)^2}{\sigma^2_j}.
\end{equation}

\subsection{Main-sequence prior}
\label{sec:msprior}

In Bayesian terms, the likelihood of obtaining the observed data $D$ given the model defined by a particular set of proposal parameters is
\begin{equation}
\mathcal{P}(D|T_0,P,\Delta F,t_T,b,M_*)\propto\exp(-\chi^2/2),
\end{equation}
but the full posterior probability distribution for the data and the model depends on the prior probability distribution for each of the model parameters. We are only interested in solutions for which the companion is a planet-sized object yielding a transit of observable depth, so we restrict the impact parameter to the range $0 < b < 1.0$, rejecting proposal steps that fall outside this range.

For most of the remaining parameters the uniform prior implied by the random-walk nature of MCMC is valid, but the stellar mass and radius are already determined {\em a priori} from the $J-H$ colour, under the implicit assumption that the star is single and on the main sequence. Under this prior assumption we expect the stellar mass to lie somewhere within an approximately gaussian distribution with mean $M_0$ and  standard deviation $\sigma_M=M_0/10$ (i.e. the same arbitrary but plausible value used to determine the average jump size in $M_*$). We use a power-law approximation to  the main-sequence mass-radius relation to define a prior probability distribution for $R_*$ with mean $R_0=M_0^{0.8}$ \citep{tingley2005} and hence standard deviation $\sigma_R=0.8(R_0/M_0)\sigma_M$.

This gives a joint prior probability distribution for the values of the proposal parameter$M_*$ and the derived physical parameter $R_*$ of the form

\begin{equation}
\mathcal{P}(M_{*,i},R_{*,i})=\exp\left(-\frac{(M_{*,i}-M_0)^2}{2\sigma^2_M}-\frac{(R_{*,i}-R_0)^2}{2\sigma^2_R}\right).
\end{equation}
Since the posterior probability distribution is $ \mathcal{P}(M_{*,i},R_{*,i})\mathcal{P}(D|T_0,P,\Delta F,t_T,b,M_*)$, we impose the prior on $M_{*,i}$ and $R_{*,i}$ by replacing $\chi^2_i$ with the logarithm of the posterior probability distribution 
\begin{equation}
Q_i \equiv \chi^2_i+\frac{(M_{*,i}-M_0)^2}{\sigma^2_M}+\frac{(R_{*,i}-R_0)^2}{\sigma^2_R}
\label{eq:msprior}
\end{equation}
as the statistic on which acceptance of a set of proposal parameters is decided.

For every new proposal set generated, the decision as to whether or not to accept the set is made via the Metropolis-Hastings rule: if $Q_i<Q_{i-1}$ the new set is accepted; if on the other hand $Q_i>Q_{i-1}$, the new set is accepted with probability $\exp(-\Delta Q/2)$, where $\Delta Q\equiv Q_i-Q_{i-1}$. The algorithm first converges to, then explores the parameter space around a constrained optimum solution that represents a compromise between fitting the light curve and reconciling the resulting stellar dimensions with prior expectations derived from the $J-H$ colour. As we shall see later, the value of the main-sequence prior $\mathcal{P}(M_{*,i},R_{*,i})$ at the global minimum of $Q$ constitutes a useful measure of how far the stellar parameters must be displaced from the main sequence values appropriate to the star's colour in order to fit the transit light curve.

We introduce a step-size controller $f$ which is adjusted on every 100th step to ensure that the acceptance rate is held close to the optimal value of 0.25 \citep{tegmark2004}. 
The value of $f$ is determined by the simple linear algorithm $f_{new}=400 f_{old} / j$, where $j$ is the total number of proposals generated and tested during the previous 100 successful steps. We find that values in the range $0.5<f<1$ achieve the necessary acceptance rate when the correct values of the parameter uncertainties are used. In any case, we find that all six parameters invariably converge to their optimal values within 500 steps. Once this ``burn-in'' phase is complete, the uncertainties on the data are  adjusted to take account of correlated errors by rescaling to ensure that the value of $\chi^2$ at the optimal solution is equal to the number of degrees of freedom.
The parameter uncertainties are subsequently allowed to evolve as the computation progresses, being recomputed from the Markov chains themselves every 100 successful steps. This ensures that the step size in each dimension samples the parameter space  adequately. After discarding the first 500 steps we compared the variances within and between five independent sub-chains, using the test statistic of \citet{gelman92}. The value of the Gelman-Rubin statistic was invariably within a fraction of one percent of unity, verifying that the chains were properly converged and well-mixed.

For all its advantages, MCMC is a discrete, stochastic process. After each MCMC run was completed, we therefore refined the values of the six proposal parameters that represent the optimum solution for $Q$ using the AMOEBA (downhill simplex) algorithm \citep{press92}, using the best-fitting step of of the Markov chain as one of the initial vertices and the first six successful steps of the chain as the others. 
The derived primary and secondary radii (in solar and jovian units), are listed alongside the best-fitting proposal parameters for all 67 stars in Table~\ref{tab:bestparms}. 

\begin{figure*}
\begin{tabular}{ll}
\psfig{figure=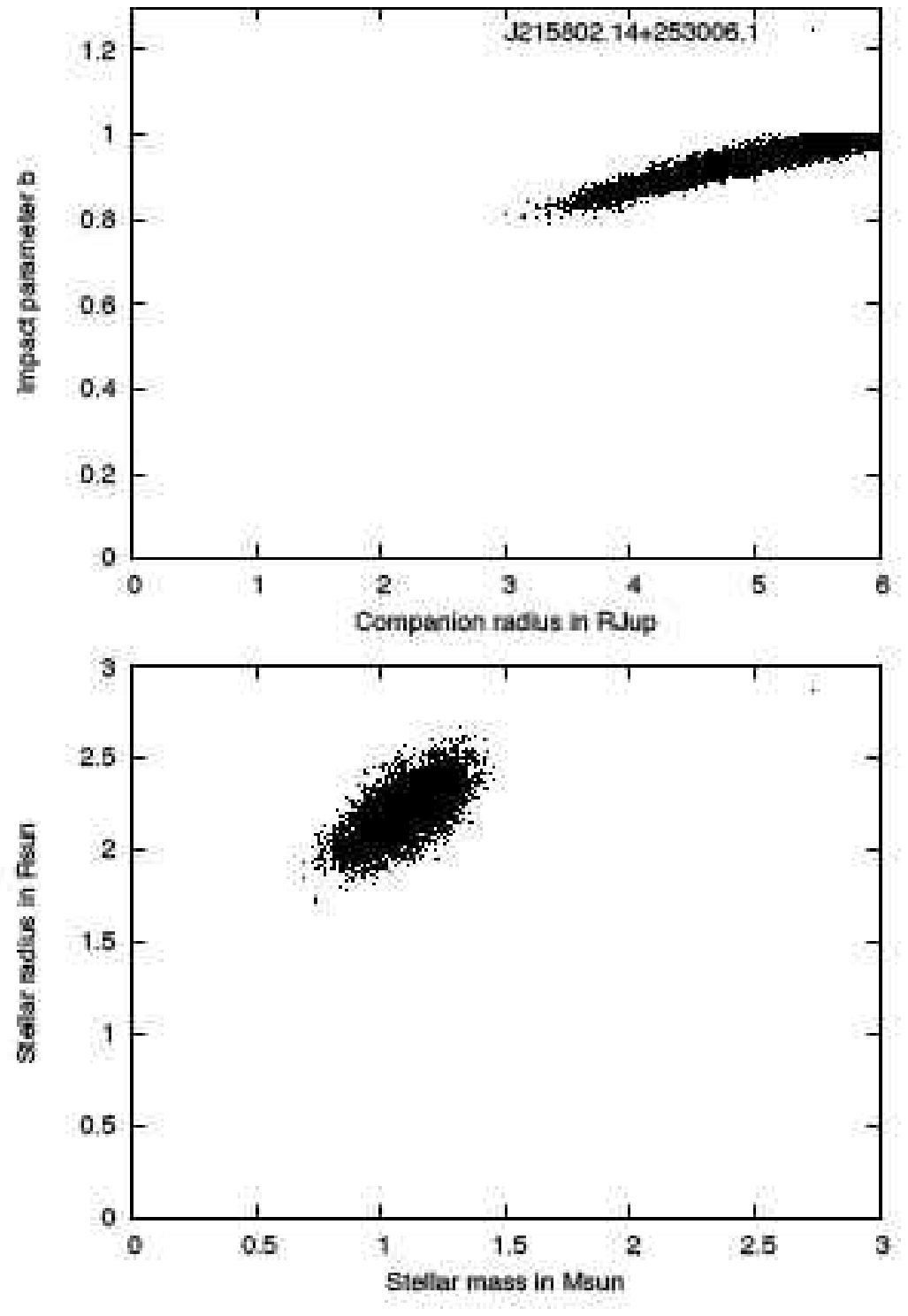,width=8.5cm}&
\psfig{figure=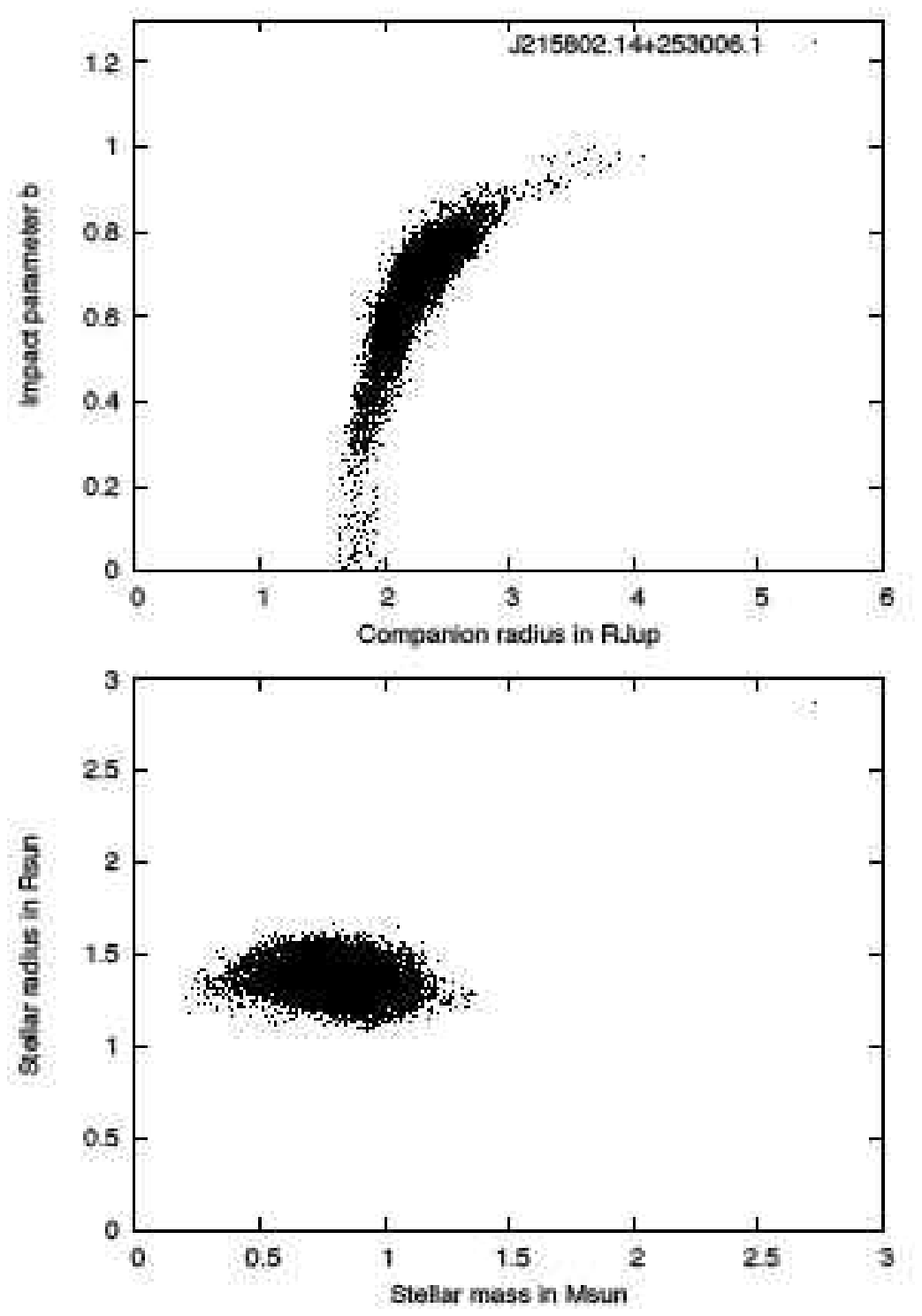,width=8.5cm}
\end{tabular}
\caption[]{Impact parameter versus planet radius and stellar radius versus stellar mass for MCMC analysis of the SuperWASP light curves of the grazing broad-lined 
spectroscopic binary 1SWASP J215802.14+253006.1. The left-hand realisation imposes a prior on the stellar mass, derived from the $J-H$ colour. The right-hand panels show the effect of imposing an additional main-sequence prior on the stellar radius.}
\label{fig:215802priors}
\end{figure*}

\subsection{Orthogonality properties}

We note that the proposal parameter set $\{T_0,P,\Delta F,t_T,b,M_*\}$ is particularly well-suited to MCMC, because the first four parameters have approximately Gaussian probability density functions that are directly related to observationally-determined quantities. As Fig.~\ref{fig:plotmatrix} illustrates, all six are close to being mutually uncorrelated. This circumvents the difficulty commonly encountered in some applications of MCMC to transit modelling, in which the normalised stellar radius $R_*/a$ is used directly in place of $t_T$. The parameters $R_p/R_*$, $R_*/a$ and $b$ are strongly covariant, leading to correlation lengths of several hundred to 1000 steps in the Markov chain \citep{holman2006,winn2007tres1} if they are used directly as proposal parameters. Shorter correlation lengths are desirable, since they allow more statistically-independent samples of the parameter space to be generated in less computing time.
In our parameter space the equivalent observational parameters $\Delta F$, $t_T$ and $b$ are only very weakly covariant. Their use reduces the correlation lengths in the chains for these parameters to between 4 and 20 steps, yet they yield $R_*/a$ and $R_p/R_*$ directly via eqs.~\ref{eq:rsa} and \ref{eq:rprs}. We note that \citet{burke2007} have arrived at a very similar conclusion, and that further shortening of the correlation length is possible if their orthogonalisation procedure is employed.

\section{Secondary  radii and impact parameters}
\label{sec:compare}

The use of the 2MASS $J-H$ colour to place prior constraints on the stellar mass and radius is of key importance in extracting essential information about the system dimensions from SuperWASP light curves. The value of the prior at the constrained optimum solution gives a powerful diagnostic for the presence of multiple stellar spectra, and for the host star's proximity to the main sequence.

\begin{figure*}
\begin{tabular}{ll}
\psfig{figure= 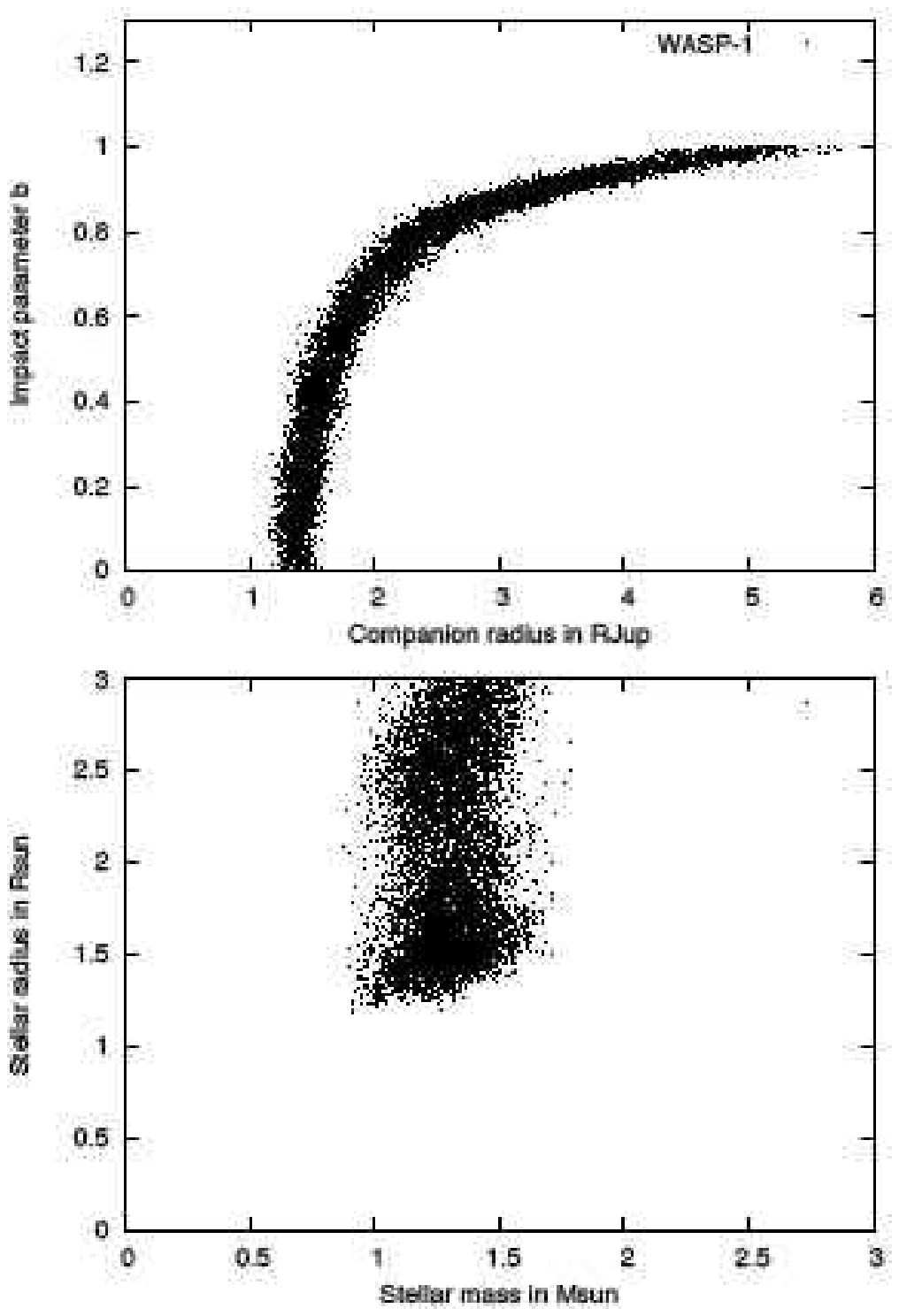,width=8.5cm}&
\psfig{figure= 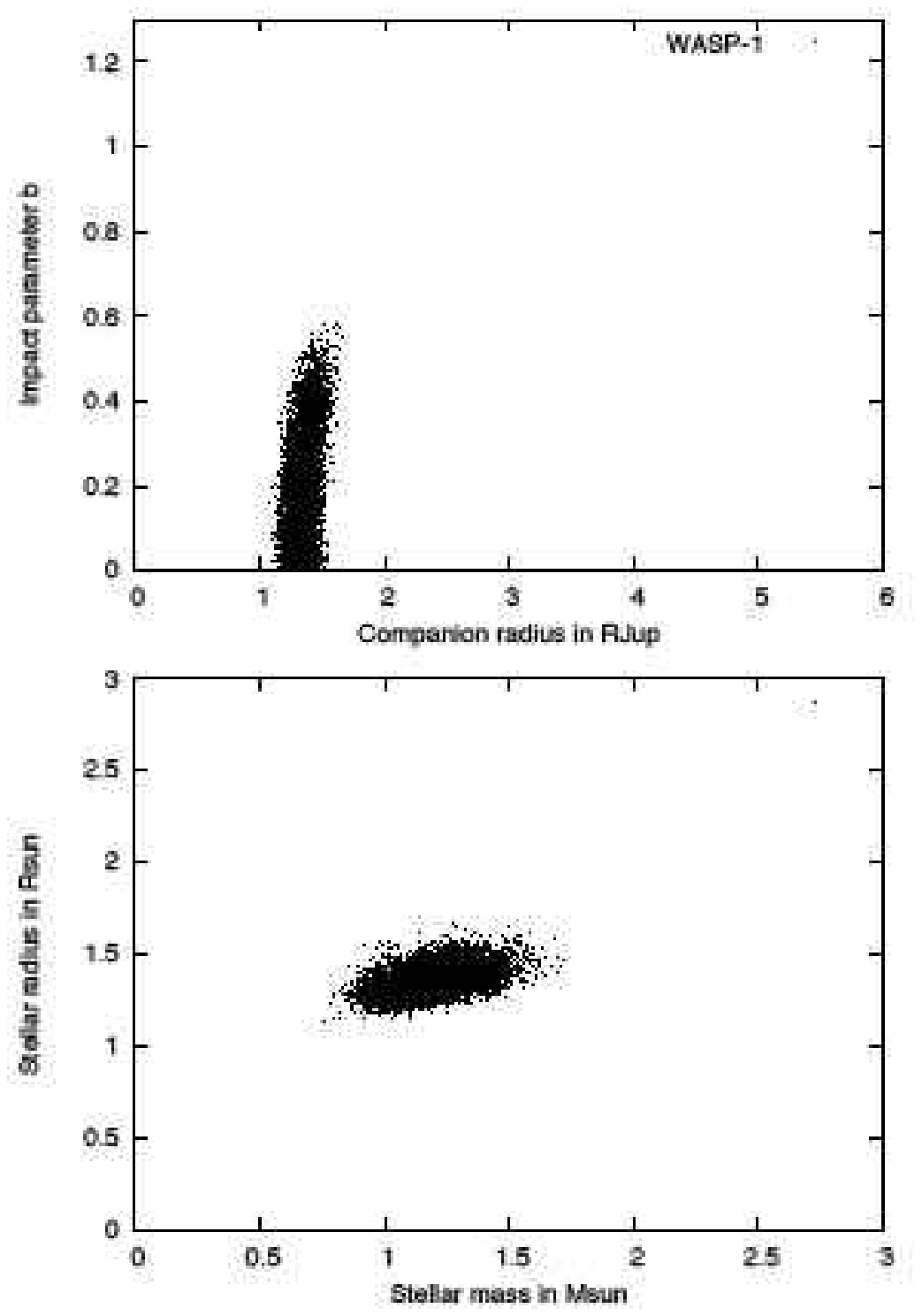,width=8.5cm}
\end{tabular}
\caption[]{As for Fig.~\ref{fig:215802priors}, for the known exoplanet host star WASP-1.}
\label{fig:wasp1priors}
\end{figure*}

\begin{figure*}
\begin{tabular}{ll}
\psfig{figure= 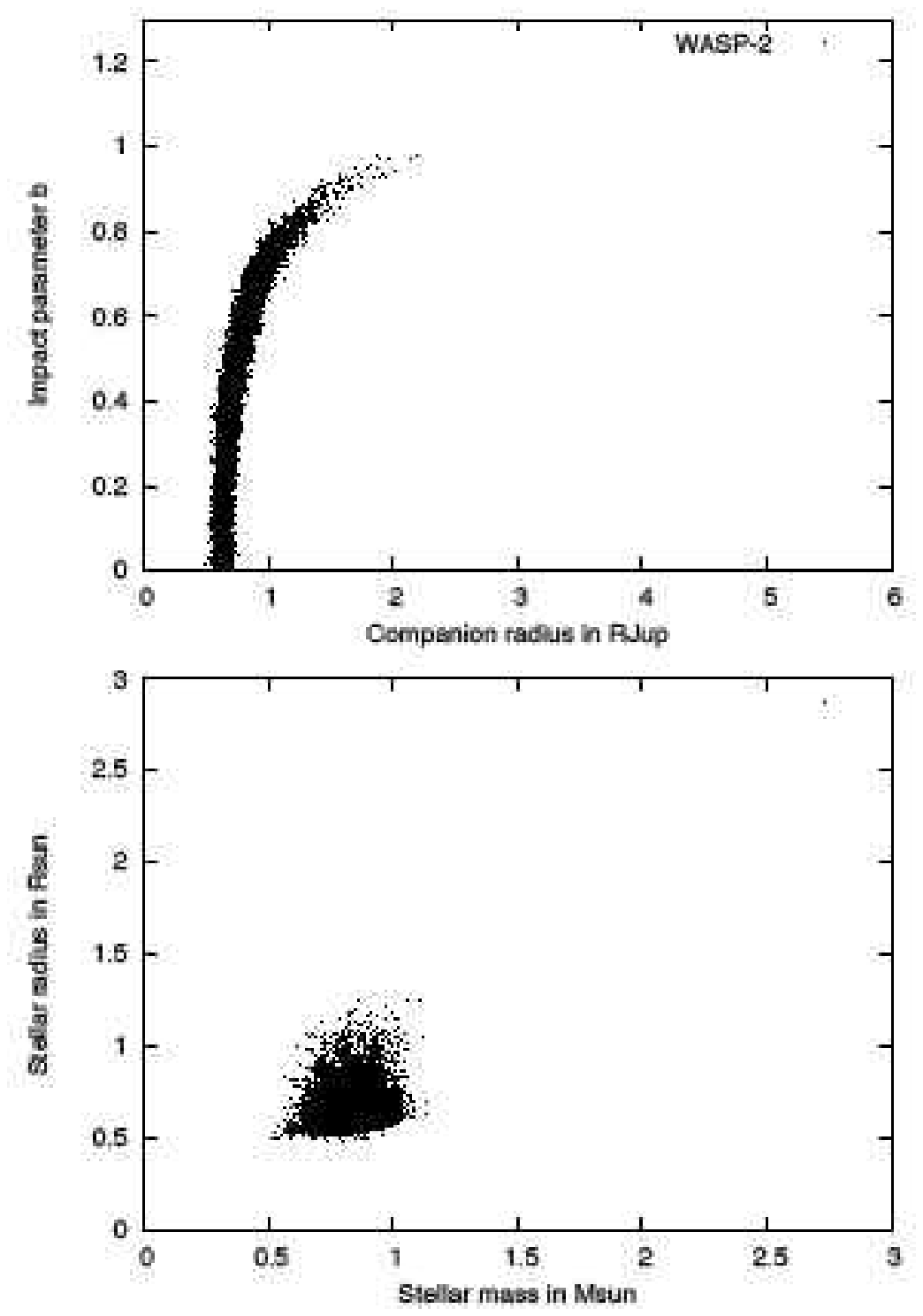,width=8.5cm}&
\psfig{figure= 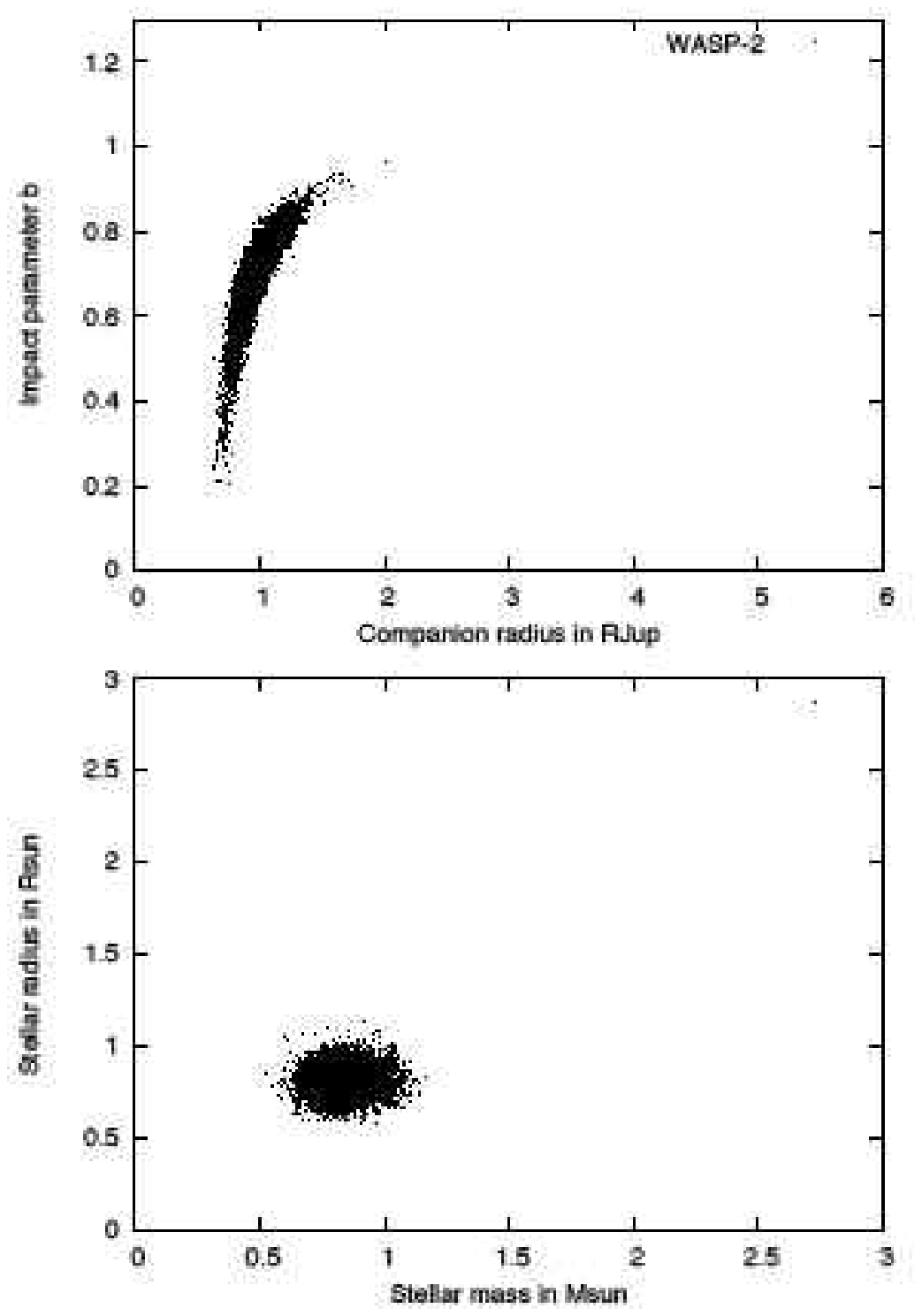,width=8.5cm}
\end{tabular}
\caption[]{As for Fig.~\ref{fig:215802priors}, for the known exoplanet host star WASP-2.}
\label{fig:wasp2priors}
\end{figure*}

To illustrate this point we show in Figs~\ref{fig:215802priors}, \ref{fig:wasp1priors} and \ref{fig:wasp2priors} the MCMC realisations of the relationship between impact parameter and planet radius for WASP-1,  WASP-2 and the broad double-lined spectroscopic binary 1SWASP J215802.14+253006.1. We  performed two MCMC analyses. The first imposed a prior on the stellar mass only. The posterior probability density is then proportional to $\exp(-Q_i/2)$, where
$$
Q_i=\chi^2_i+\frac{(M_*,i-M_0)^2}{\sigma^2_M}
$$
is used as the statistic to which the Metropolis-Hastings rule is applied. The second analysis imposed the full main-sequence prior as described in Eq.~\ref{eq:msprior}.

\begin{table}
\caption[]{Comparison of stellar and planetary dimensions derived by other authors from 
high-precision photometry with those derived from SuperWASP photometry in the present study using the constrained MCMC analysis. The middle columns show
the radius values of Shporer et al and Charbonneau et al. rescaled as $M_*^{1/3}$ to account for the differences in the other authors' assumed and our fitted stellar masses. }
\label{tab:compare}
\begin{tabular}{llll}
\hline\\
Parameter & Shporer et al & Charbonneau et al& This study \\
                  & (2007) rescaled  & (2007) rescaled & \\ 
\hline\\
WASP-1: & & & \\
$R_*/R_\odot$ & $1.46\pm 0.06$ & $1.490\pm 0.033$ & $1.382^{+0.047}_{-0.116}$ \\
$R_p/\rmsub{R}{Jup}$ & $1.44\pm 0.06$ & $1.480\pm 0.040$ & $1.358^{+ 0.048}_{-0.104}$ \\
$b$ &$0.03\pm 0.17 $ & $< 0.336$ & $0.215^{+ 0.091}_{-0.156}$ \\
$M_*/M_\odot$ & 1.24 (matched) & 1.24 (matched) & $1.24^{0.12}_{-0.17}$ \\
& & & \\
WASP-2: & & & \\
$R_*/R_\odot$ & -- & $0.830\pm 0.033$ & $0.834^{+0.066}_{-0.083}$ \\
$R_p/\rmsub{R}{Jup}$ & -- & $1.059\pm  0.051$ & $1.017^{+ 0.111}_{-0.153}$ \\
$b$ & -- & $ 0.731\pm0.026$ & $0.753^{+ 0.033}_{-0.151}$ \\
$M_*/M_\odot$ & -- & 0.84 (matched) & $0.84^{0.07}_{-0.11}$ \\
\hline\\
\end{tabular}
\end{table}

1SWASP J215802.14+253006.1 was selected as a candidate, having been found to exhibit transits 0.024 magnitudes deep. Its 2MASS $J-H=0.29$ suggests an early G spectral type. It passed the transit depth and duration tests of \citet{tingley2005} and all other photometric selection tests. A single SOPHIE spectrum, however, revealed the presence of two rotationally-broadened stellar spectra. 

The left-hand panel of Fig.~\ref{fig:215802priors} shows the MCMC joint posterior probability distributions of the impact parameter and secondary radius, and of the primary mass and radius of 1SWASP J215802.14+253006.1. The MCMC results for which the prior is applied to the the stellar mass only show that the impact parameter is high, indicating a grazing eclipse. The mass-radius plot reveals a further inconsistency: the inferred primary radius is twice that expected for a main-sequence star. Imposing the further Bayesian constraint on the primary radius does little to improve matters. The prior's attempt to pull the primary's dimensions toward the main sequence is thwarted by the need to fit the V shape of the transit light curve. The impact parameter is reduced somewhat, but the primary mass is pulled toward a value too low to be consistent with the colour. The primary's radius remains too large for its mass. The probability that the secondary has a radius less than 1.5 times that of Jupiter remains negligible, while the statistic representing the main-sequence prior is driven to the very high value $S=32.53$. The star 1SWASP J215802.14+253006.1 thus fails two important statistical tests, and can be eliminated as a viable exoplanet transit candidate without the need for spectroscopic follow-up.


For those systems that appear to have single main-sequence primaries, the mass-radius constraint also serves to reduce the uncertainty in the impact parameter of the transit. By imposing a prior that pulls the stellar radius toward the main-sequence mass-radius relation, we reduce the degeneracy between the stellar radius and impact parameter that arises from the transit duration. Since the ratio of the secondary's radius to that of the primary depends on the transit depth and impact parameter, the uncertainty in the secondary's radius is substantially improved.

In Table~\ref{tab:compare} we compare the stellar and planetary dimensions inferred from the SuperWASP light curves using this method, with the values determined by \citet{shporer2007} from high-quality $I$-band transit photometry of WASP-1, and by \citet{charbonneau2007} from high-quality $z$-band observations of both WASP-1 and WASP-2. The imposition of the main-sequence prior fixes the impact parameter at values for both stars that agree well with those of both sets of authors. Our values for the radii of both planet and host star in the WASP-2 system are also in excellent agreement with those of Charbonneau et al. 

Our radii for WASP-1 and its planet are systematically smaller than those found by Charbonneau et al and by Shporer et al. These authors both determine a primary radius that is consistent with the star being somewhat above the main sequence and hence slightly evolved, whereas our method artificially pulls the radius toward the expected main-sequence value. The mass that we derive from $J-H$ is consistent with the range of values determined by \citet{stempels2007} from their recent spectroscopic abundance analysis of WASP-1. The inflated radius found by Shporer et al. and Charbonneau et al. is consistent with the rather low $\log g = 4.28\pm 0.15$ found by Stempels et al. WASP-1 thus provides valuable verification that the selection criteria  suggested in Section~\ref{sec:criteria} below are not so strict as  inadvertently to exclude inflated planets orbiting slightly-evolved stars.

\section{Selection of exoplanetary transit candidates}
\label{sec:criteria}

Instruments such as SuperWASP are capable of detecting the transits of gas-giant planets orbiting stars on or at most slightly above the main sequence. The $J-H$ colour of a viable candidate should thus yield a stellar mass and radius that lie near the main sequence, and which are consistent with the duration of the transit. The value of the main-sequence prior $\mathcal{P}(M_{*},R_{*})$ measures the displacement of the fitted stellar mass and radius at the global minimum of $Q$ from the main sequence values we expect from the star's $J-H$ colour. 

The radius derived for the companion should be consistent with its being a planet. Although the SuperWASP light curves are not always of sufficient quality to discern the shape of the transit, the main-sequence prior helps to break the well-known degeneracy between the impact parameter $b$ and the stellar radius $R_*$ (which is strongly correlated with the companion radius via the transit depth). The use of MCMC allows us to determine directly the probability that the companion radius is less than some specified amount.

For convenience we define the quantity
$$
S = -2\ln \mathcal{P}(M_*,R_*) = \frac{(M_{*,i}-M_0)^2}{\sigma^2_M}+\frac{(R_{*,i}-R_0)^2}{\sigma^2_R}
$$ 
as a measure of the discrepancy between the constrained optimum stellar dimensions and the values derived from $J-H$.

\begin{figure*}
\begin{tabular}{ll}
\psfig{figure=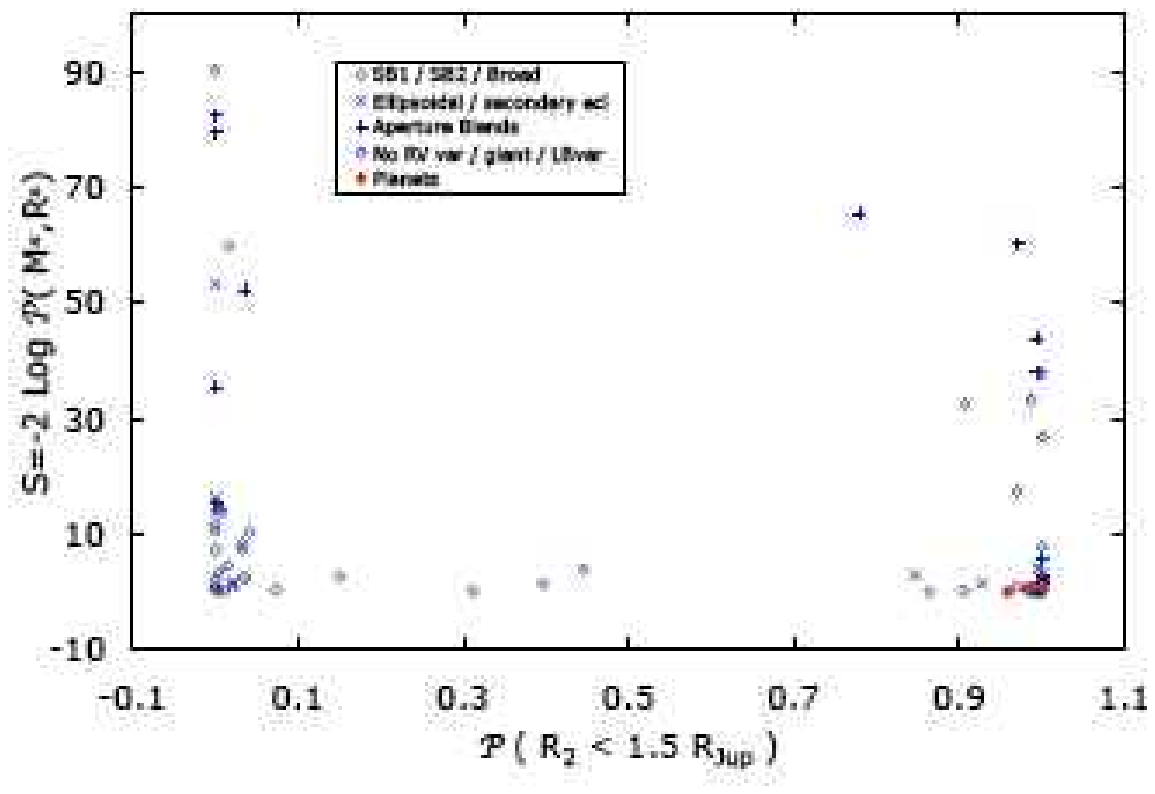,width=8.5cm}&
\psfig{figure=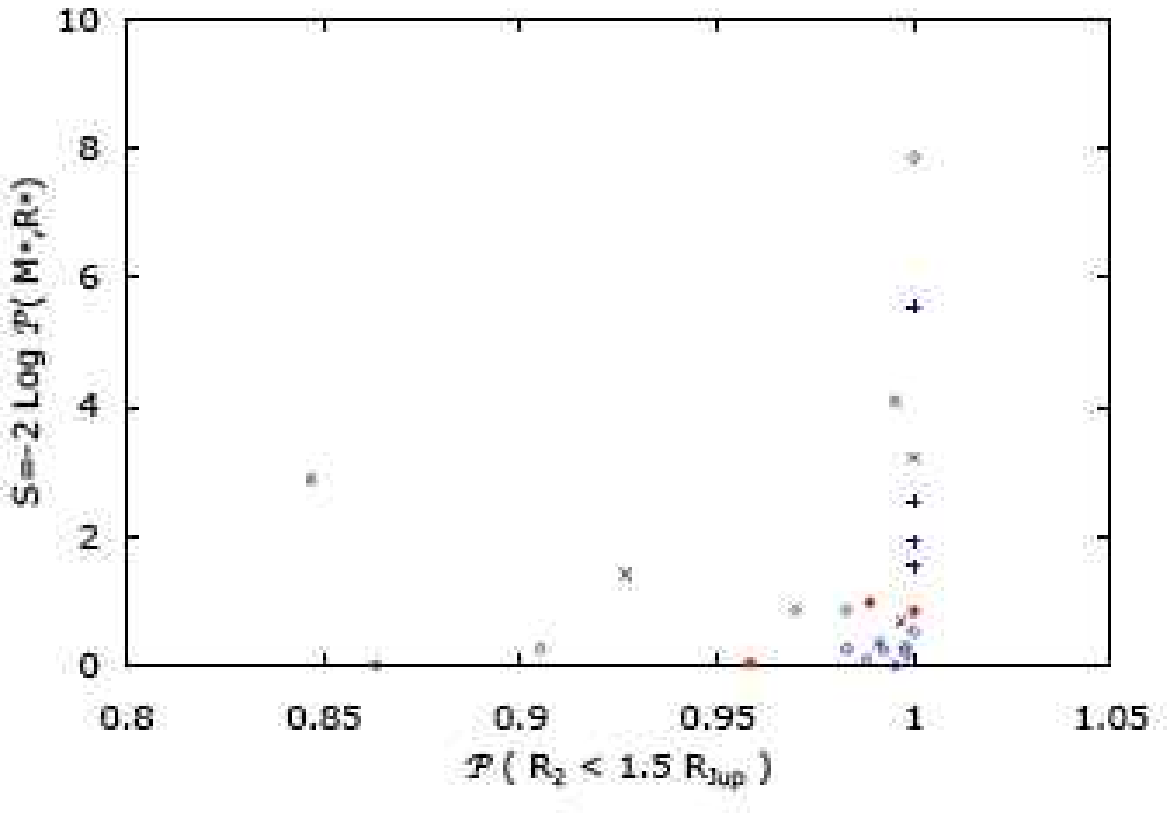,width=8.5cm}
\end{tabular}
\caption[]{Logarithm of main-sequence prior probability versus probability that companion radius is less than $1.5 \rmsub{R}{Jup}$ for known planets (red circles) and other forms of astrophysical false positive. Open circles and diamonds represent spectroscopic binaries and unresolved aperture blends, which could only be identified spectroscopically as false positives. Upright and diagonal crosses are eclipsing binaries and resolved aperture blends, detected using the methodology of Section~\ref{sec:sececl} and \ref{sec:apblend} respectively. The right-hand panel gives an enlarged view of the lower-right corner of the left panel.}
\label{fig:prp_dchisqmr}
\end{figure*}

In Fig.~\ref{fig:prp_dchisqmr} we plot $S$ against the probability that the companion has a radius less than 1.5 times that of Jupiter, for the three planet-host stars and the various classes of astrophysical false-positive in our survey sample. The diagram shows clearly that the sample is cleanly partitioned into objects with high and low probabilities of having Jupiter-sized secondary components.

There is a tight cluster of objects in the lower-right corner of Fig.~\ref{fig:prp_dchisqmr}, with the main-sequence prior giving $S < 1$ and probabilities greater than 0.95 that the secondary radius is less than 1.5 times that of Jupiter. This cluster includes all three known planets in the sample, together with two unresolved aperture blends and ten spectroscopic binaries, two of which can be eliminated photometrically as exhibiting ellipsoidal variations or unequal odd- and even-numbered transits.

In selecting targets for spectroscopic followup it would be unwise to set thresholds as tight as the bounds of this cluster. The light curve of the recently-discovered TrES-3 \citep{odonovan2007} is a good illustration of the need for caution. The light curve of TrES-3  was recorded by the SuperWASP cameras as 1SWASP J175207.01+373246.3 and its transits were detected with the correct period \citep{lister2007} even though it did not satisfy our original selection criteria for followup at high priority. Applying the MCMC test to the SuperWASP light curve of TrES-3, we find that $S=0.06$, indicating a stellar mass and radius consistent with expectations, but $\mathcal{P}(R_2<1.5\rmsub{R}{Jup})=0.236$.  The reason for the apparently low probability of the secondary having a planet-like radius arises because TrES-3 has a high impact parameter. Limb darkening produces a strong positive correlation between $b$ and $R_p$, giving a range of solutions with  $0.8<b<1$ and $1.3 < R_p< 2.0 \rmsub{R}{Jup}$. By requiring candidates for followup spectroscopy to have $S<5$ and $\mathcal{P}(R_2<1.5\rmsub{R}{Jup})>0.2$, we leave a sufficient margin to allow for the possibility of detecting inflated planets transiting slightly evolved main-sequence stars at high impact parameters. 

Six of the 19 binaries that satisfy these more relaxed selection criteria are represented by diagonal crosses in Fig.~\ref{fig:prp_dchisqmr}, having been identified photometrically as mimics from their ellipsoidal variations or secondary eclipses. Similarly, three of the five surviving blends, denoted by upright crosses in Fig.~\ref{fig:prp_dchisqmr}), fail the photometric aperture-blend test. They could thus have been eliminated without requiring spectroscopic follow-up. With the wisdom of hindsight, we would thus retain 18 candidates from our original sample of 67 stars for spectroscopic followup, of which two are unresolved aperture blends,  thirteen are stellar binaries and three are planet host stars. The final selection decisions are given in the last column of Table~\ref{tab:classify}.

\section{Dwarf-Giant separation}
\label{sec:rpm}

If proper motion information is available, a useful secondary check can be made on the luminosity class of the primary. Evolved giant and sub-giant stars have very similar photometric colours to their main-sequence counterparts and are frequently a source of contamination in exoplanetary transit surveys. They can easily be identified, however, from their proper motions. Giant stars are significantly more distant than dwarfs of the same magnitude, and so exhibit substantially smaller proper motions.

\begin{figure}
\begin{center}
\psfig{figure=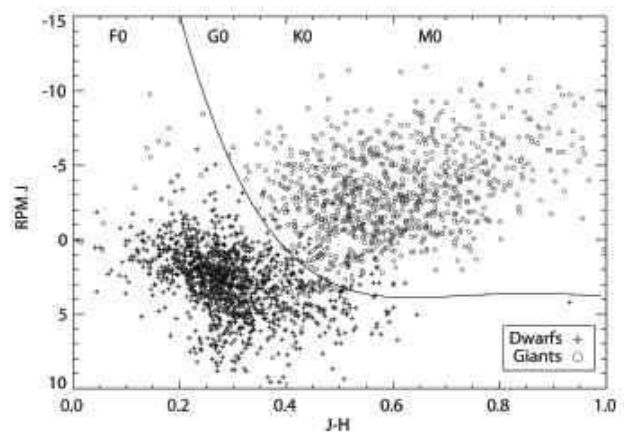,width=8.5cm}
\caption[]{Reduced proper motion against $J-H$ for dwarfs (log g $\ge$ 4.0) and giants (log $\le$ 3.0).}
\label{fig:rpmj}
\end{center}
\end{figure}

\begin{figure}
\begin{center}
\psfig{figure=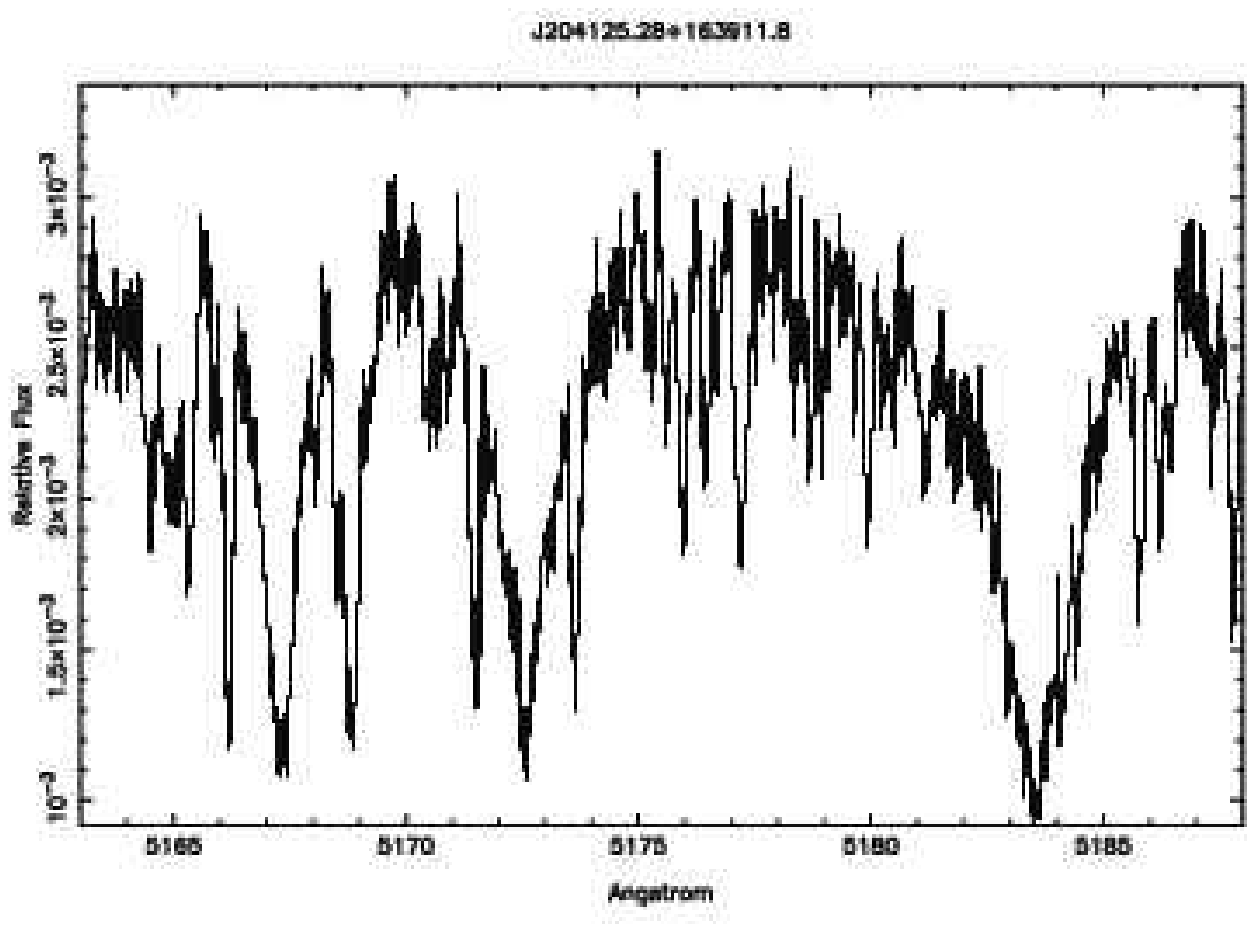,width=8.5cm}
\psfig{figure=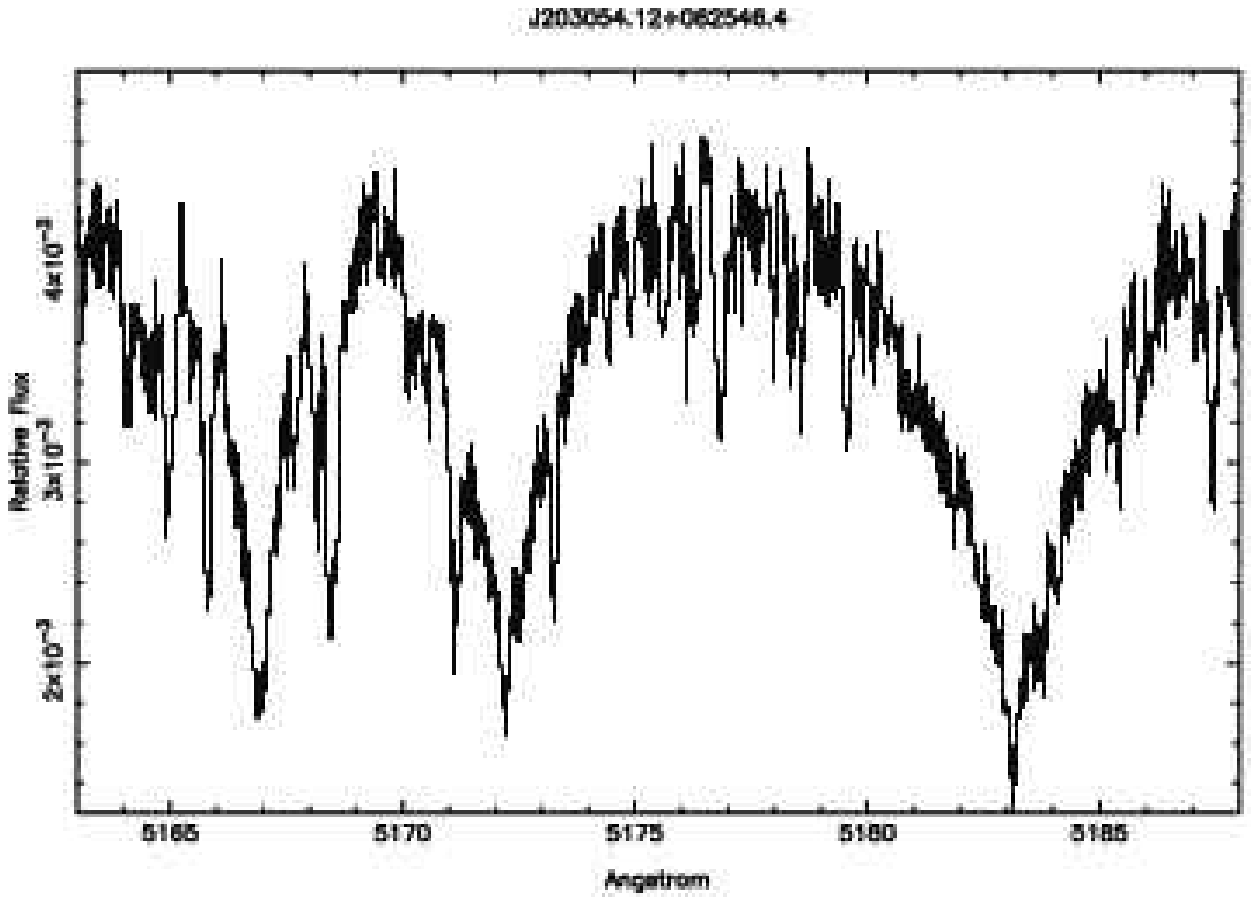,width=8.5cm}
\caption[]{SOPHIE spectra of the Mg I b triplet region in the K giant 1SWASP J204125.28+163911.8 (upper) and the K dwarf planet-host star WASP-2. Note the difference in pressure broadening of the wings of the Mg Ib triplet lines at 5167.3\AA, 5172.7\AA and 5183.6\AA.}
\label{fig:mgwings}
\end{center}
\end{figure}

\citet{gould2003} found that a plot of Reduced Proper Motion (RPM) against Tycho-2 $B_T-V_T$ provided an effective means of separating giants from dwarfs. We adapted this method using the $J-H$ colour index and proper motions from the all-sky USNO-B1.0 catalogue \citep{monet2003}. Our model is calibrated using stars from the catalogues of \citet{valenti2005} and \citet{cayrel2001} which list over 2000 FGK stars with high resolution, high signal-to-noise spectra from which the surface gravity has been determined to better than $\pm$ 0.1dex. The stars were cross-matched with the USNO-B1.0 catalogue to determine proper motions and with the 2MASS catalogue to determine $J-H$ using a simple nearest-neighbour search within a radius of 10 arcsec. The stars were partitioned into dwarfs with $\log g \ge 4.0$ and giants with $\log g \le 3.0$. These conservative limits reduce the chances of accidental elimination of mildly evolved dwarfs. This resulted in 1526 dwarfs and 1145 giants. Although this is a much smaller sample than that used by \citet{gould2003}, the luminosity class determination is based on more accurate spectroscopic measurements rather than broad-band photometric calibrations.
The Reduced Proper Motion, ${\rm{H_J}}$, is defined as:
\begin{equation}
{\rm{H_J}} = J + 5\log(\mu)
\end{equation}
where $J$ is the 2MASS $J$ apparent magnitude and $\mu$ is the proper motion. In Figure~\ref{fig:rpmj} we plot ${\rm{H_J}}$ against $J-H$. A clean separation is seen between the giants and dwarfs. As predicted the giants, in general, have lower Reduced Proper Motions than the dwarfs with the same $J-H$ colour. A 4th order polynomial has been fitted to a subjectively-drawn boundary showing the best separation between the two luminosity classes where:

\begin{eqnarray*}
y &=& -141.25(J-H)^4 + 473.18(J-H)^3 \\
&& -583.6(J-H)^2 + 313.42(J-H) -58
\end{eqnarray*}

The greatest cross-contamination occurs for early K-dwarfs/giants ($J-H \sim 0.5$) in agreement with Gould and Morgan's findings. Unfortunately, this sample lacks large numbers of K and M dwarfs due to the inherent difficulty in obtaining high signal-to-noise spectra for these faint stars. As a result, the dwarf/giant determination becomes increasingly uncertain with decreasing temperature.

We stress that although the reduced proper motion test is a very useful secondary check on the MCMC statistics, we do not use it as a primary selection criterion.  It carries too high a risk of inadvertently rejecting genuine K dwarfs which just happen to have low proper motions. Nonetheless, we note that all the objects classified as giants on the basis of their reduced proper motions fail one or other of the selection criteria. In particular, we note from Table~\ref{tab:classify} that the photometric aperture blends are almost invariably also classified as giants. The RPM giants among the aperture that were observed with SOPHIE showed the low pressure-broadening in the line wings expected of giants (Fig.~\ref{fig:mgwings}).

\section{Discussion and conclusions}

We conclude that MCMC analyses of transit light curves can be carried out in such a way as to break the degeneracy between impact parameter and stellar radius, if a Bayesian prior is imposed to constrain the stellar mass and radius to values consistent with the host star's $J-H$ colour. The posterior probability distributions for the stellar and planetary parameters are obtained directly from the Markov chain. They provide a straightforward estimate of the probability that the secondary has a radius consistent with planetary status.

The prior carries the implicit assumption that the star is single and on the main sequence. We have found in this study that in a large fraction of astrophysical false positives for which these assumptions are invalid, the value of the prior at the optimum solution (in the sense of minimising the quantity $Q_i$ defined in Eq.~\ref{eq:msprior}) is significantly greater than unity, if the uncertainty in the stellar mass derived from $J-H$ is fixed (arbitrarily but reasonably) at 10 percent. The three known planets in our sample and the newly-discovered TrES-3 all yield values of the prior less than unity at their optimal solutions.

The probability that the secondary is of planetary dimensions and the value of the prior at the optimum solution provide the two-dimensional classification scheme shown in Fig.~\ref{fig:prp_dchisqmr}. For the sample of stars studied here, which had already passed the preliminary selection procedures of \citet{cameron2006}, this classification scheme is effective at identifying 75 percent of all types of aperture blend. The figure improves to 90 percent if the flux ratio in different photometric apertures is used to identify  blends of non-variable stars with eclipsing binaries near the edge of the photometric aperture.

The Bayesian selection scheme also eliminates roughly 67 percent of all spectroscopic binaries. It is particularly effective at removing grazing stellar binaries, most of which yield very low probabilities of having planet-sized secondaries, and many of which also produce values of the main-sequence prior that are too high to be consistent with a single, main-sequence primary of the observed $J-H$ colour. Grazing binaries frequently show either ellipsoidal variations or unequal eclipses when folded on twice the best period found by the accelerated box least-squares algorithm, giving an effective second line of defence against this type of impostor.

The remaining candidates have demonstrably low companion radii, generally low impact parameters, and primaries that appear to be close to the part of the main sequence suggested by their observed $J-H$ colours. Many of them are likely to be astrophysically interesting binaries, whose Jupiter-sized stellar or substellar companions will add to our knowledge of the mass-radius relation at and below the bottom of the main sequence once their spectroscopic orbits are established. Other teams pursuing similar studies have already achieved important advances in this area \citep{bouchy2005,pont2005a,pont2005b,pont2006}. In planet-hunting terms alone, however, the prospects for a high ``hit rate'' in future follow-up campaigns are good: in our sample, confirmed transiting planets comprise nearly 20 percent of the survivors. 

\section*{Acknowledgments}

This paper is based in part on observations made at Observatoire de Haute Provence (CNRS), France; with the 2-m-Alfred Jensch telescope of the Th\"uringer Landessternwarte; and with the Isaac Newton Telescope, operated on the island of La Palma by the Isaac Newton Group in the Spanish Observatorio del Roque de los Muchachos of the Instituto de Astrof'sica de Canarias. The WASP project is funded and operated by Queen's University Belfast, the Universities of Keele, St Andrews and Leicester, the Open University, the Isaac Newton Group, the Instituto de Astrofisica de Canarias, the South African Astronomical Observatory and by PPARC. This publication makes use of data products from the Two Micron All Sky Survey, which is a joint project of the University of Massachusetts and the Infrared Processing and Analysis Center/California Institute of Technology, funded by the National Aeronautics and Space Administration and the National Science Foundation. This research has made use of  the VizieR catalogue access tool, CDS, Strasbourg, France. ME acknowledges financial support by the DLR grant No. 50 OW 0204. We thank the anonymous referee for helpful comments that led to several improvements in the manuscript.

\begin{figure*}
\begin{center}
\begin{tabular}{ll}
\psfig{figure=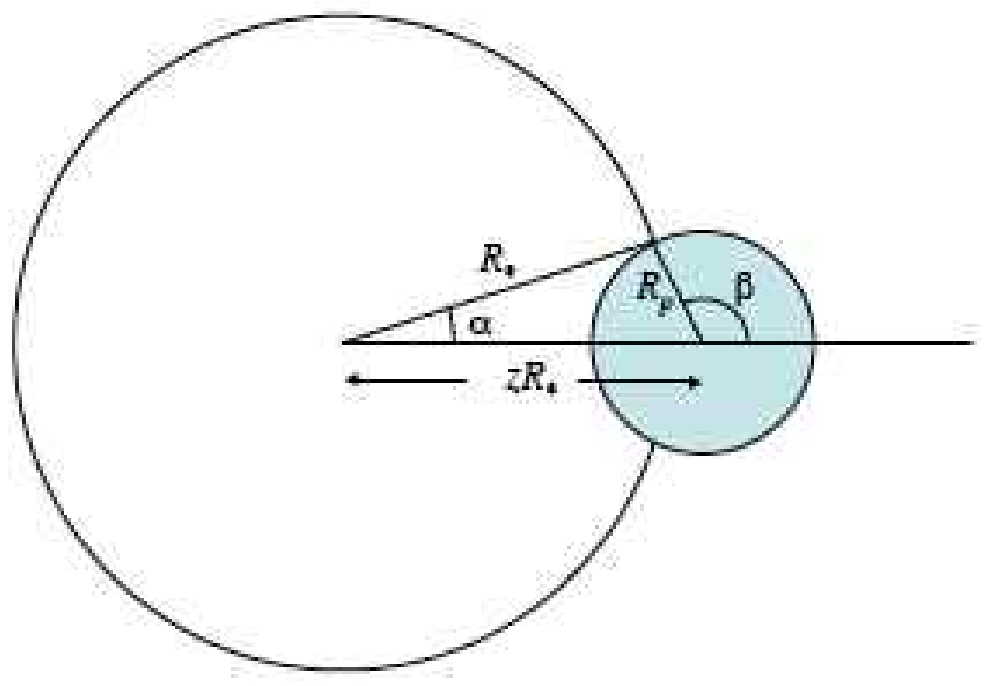,width=8.5cm}
\psfig{figure=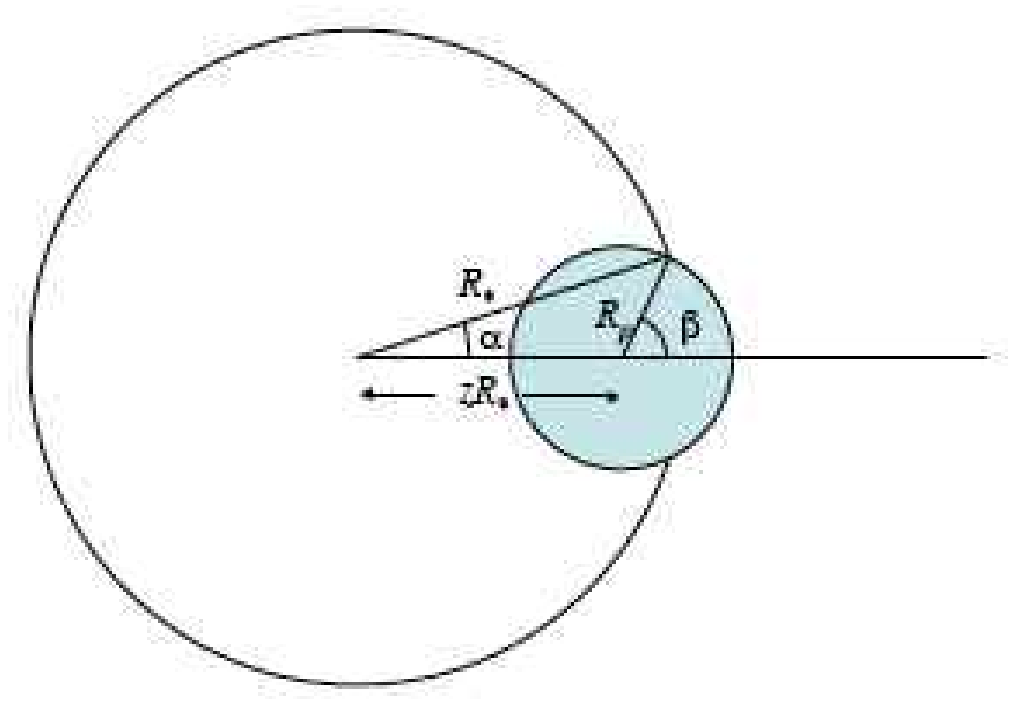,width=8.5cm}
\end{tabular}
\caption[]{Geometry of partial transit phases, showing the angles $\alpha$ and $\beta$ used in computing the fraction of the stellar disc obscured by the companion.}
\label{fig:triangles}
\end{center}
\end{figure*}

\bibliographystyle{mn2e}

\appendix
\section{Observable and physical parameters}
\label{sec:transform}

\citet{seager2003} characterised the light curve of a transit candidate in terms of four observable quantities: the transit depth $\Delta F$ expressed as a fraction of the system flux outside transit, the total transit duration $t_T$ from first to fourth contact, the duration $t_F$ of the flat part of the transit from second to third contact, and the orbital period $P$. Assuming that the transit is total, that all the light emanated from the primary star, and that the orbit was circular, they derived expressions for the ratio of the planet's radius $R_p$ to that of the star $R_*$, the ratio of $R_*$ to the orbital separation $a$, and the impact parameter $b=a \cos i/R_*$. Kepler's third law  then gives the stellar density $\rho/\rho_\odot$, under the further assumption that the mass ratio $M_p/M_*<<1$. 

Here we adopt a similar approach, generalised to allow for grazing as well as total eclipses. Following \citet{seager2003} and using the approximation that $R_p+R_* << a$ we determine the stellar radius from the transit depth, duration and impact parameter via the relation:
\begin{equation}
\frac{R_*}{a}=\frac{t_T}{P}\frac{\pi}{(1+\sqrt{\Delta F})^2-b^2}.
\label{eq:rsa}
\end{equation}
For transits where the companion is fully silhouetted against the primary during the middle part of the transit we neglect limb darkening and define
\begin{equation}
\frac{R_p}{R_*}=\sqrt{\Delta F}.
\label{eq:rprs}
\end{equation}
For partial transits, or at phases where the projected separation z of the centres lies
in the range $1-R_p/R_* < z < 1+R_p/R_*$, the fraction of the primary's disc that is
obscured by the companion is
\begin{equation}
\Delta F = \frac{R_p}{R_*}\left(\pi-\beta+\sin\beta\cos\beta+\alpha-\sin\alpha\cos\alpha\right).
\end{equation}
The angle $\alpha$ is subtended at the centre of the primary by the line of centres 
and the radius of the primary at the point where the two limbs intersect. The angle $\beta$ is subtended at the centre of the planet by the line of centres and the radius of the planet at the point where the two limbs intersect (Fig.~\ref{fig:triangles}). The angles $\alpha$ and $\beta$ are defined by
\begin{equation}
\cos\alpha=\frac{3}{2}z+\frac{1+(R_p/R_*)^2}{2z}
\end{equation}
and
\begin{equation}
\cos\beta = \frac{1-(R_p/R_*)^2-z^2}{2zR_p/R_*}.
\end{equation}

For cases where $R_p/R_*< 0.3$ or so, the fractional obscuration of the primary's disc
can be approximated to better than 4 percent by a linear function of the separation $z$ of the centres:
\begin{equation}
\Delta F = \frac{R_p}{2R_*}\left(1+ \frac{R_p}{2R_*}-z\right).
\end{equation}
This equation is quadratic in $R_p/R_*$ and has the positive solution
\begin{equation}
\frac{R_p}{R_*}=\frac{z-1+\sqrt{(1-z)^2+8\Delta F}}{2}.
\end{equation}
Since  $z=b$ at mid-transit, the ratio of the planetary and stellar radii can be determined from the transit depth $\Delta F$ at mid-transit even for grazing eclipses, provided $b$ is known. 

In practice, the combination of observational errors and the effects of limb darkening make it very difficult to determine the duration $t_F$ of  the middle phase of the transit with any degree of reliability. For this reason we parametrise the form of the transit profile in terms of $\Delta F$, $t_T$, $b$ and $P$. Like \citet{seager2003} we use Kepler's third law to determine the stellar density in solar units:
\begin{equation}
\frac{\rho}{\rho_\odot}=0.0134063\left(\frac{a}{R_*}\right)\left(\frac{P}{1\mbox{day}}\right)^{-2}.
\end{equation}

In order to close the system, we need either an additional equation or an independent determination of the stellar mass. It is then straightforward to determine the stellar radius (in solar units) from the stellar density.

\citet{seager2003} chose the former approach, imposing a main-sequence mass-radius relation on the primary of the form $R_*\propto M_*^{0.8}$. We choose instead to use the $J-H$ colour index taken from the 2MASS Point Source Catalogue of \citet{cutri2003} to estimate the stellar effective temperature and hence the corresponding main-sequence stellar mass, as described in Appendix \ref{sec:jhcalib}.

\section{Radius estimation using $J-H$}
\label{sec:jhcalib}

\begin{figure}
\begin{center}
\psfig{figure=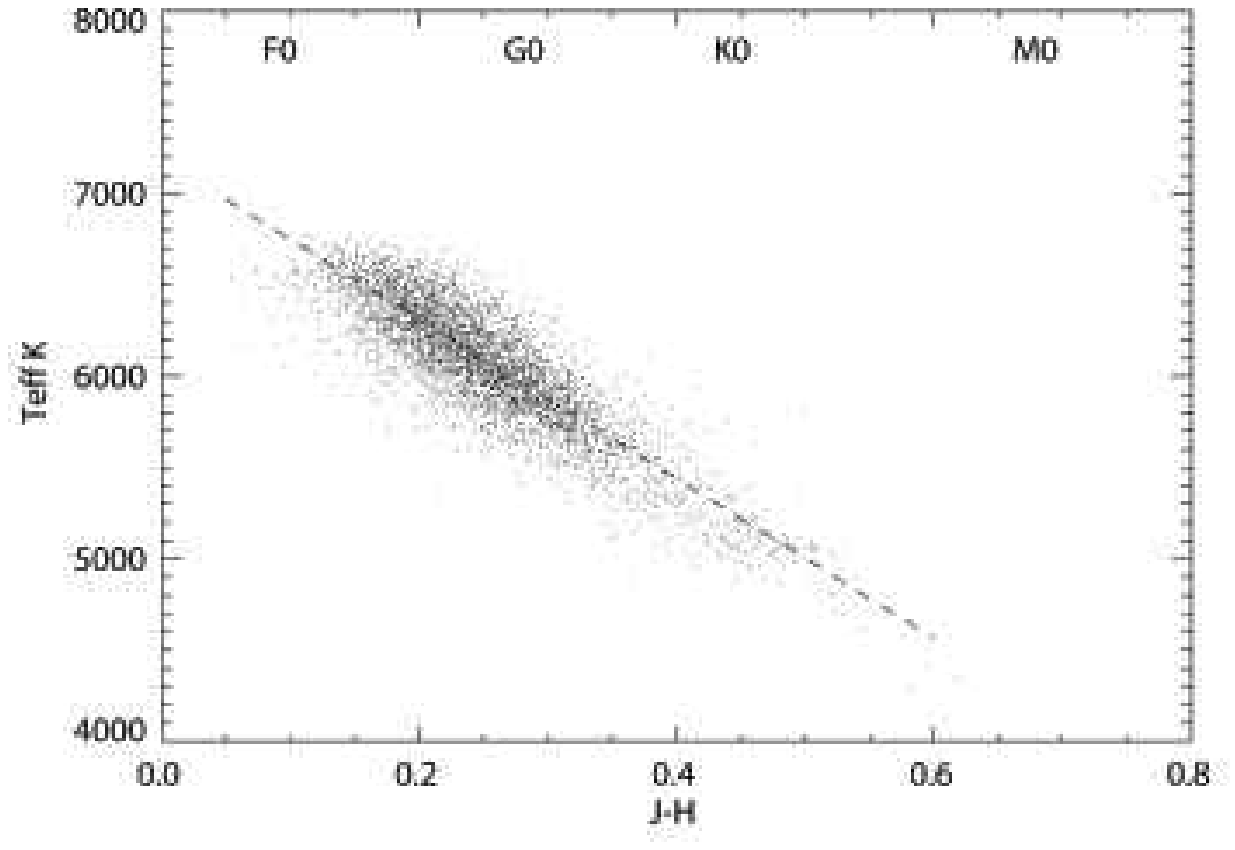,width=8.5cm}
\caption[]{$J-H$ colour index against temperature for FGK dwarfs from the Tycho-2 catalogue (only 1/15 datapoints are shown for clarity).}
\label{fig:jh}
\end{center}
\end{figure}

The 2MASS $J-H$ colour index was calibrated using stellar temperatures from a sample of 100,000 Tycho-2 FGK dwarf stars from \citet{ammons2006}, selected to assist in target selection for radial-velocity planet searches. The temperatures were calculated by fitting spline functions to broad-band Hipparcos/Tycho-2 and 2MASS photometry using a training set of stars observed with the Keck HIRES (High Resolution Echelle Spectrometer). The temperature model is quoted with an accuracy of better than 100K.
We extracted a subset of approximately 65000 stars from this sample for which the errors in the effective temperature were less than 150K and the photometry errors less than 1 percent. The $J-H$ index was then plotted against temperature (Fig.~\ref{fig:jh}) and fitted with a linear relation
\begin{equation}
\Teff = -4369.5(J-H) + 7188.2,
\label{eq:jh_teff}
\end{equation}
which is valid over the approximate temperature range $4000 < \rmsub{T}{eff} < 7000$ K.  The $J-H$ index is preferred over $J-K$ as the 2MASS errors for $H$ magnitude are generally lower than for $K$.

The scatter of the data about the linear fit is 114K and is an improvement of the fit given by the relations of \citet{cox2000}.

The stellar radius is then calculated from a polynomial fit to the temperature/radius relation for main-sequence stars tabulated in Appendix B1 of \citet{gray1992}, which is again valid over the temperature range  $4000 < \rmsub{T}{eff} < 7000$ K.
\begin{eqnarray*}
\frac{R_*}{R_\odot} &=& -3.925\times 10^{-14}(\Teff)^4 + 8.3909\times 10^{-10}(\Teff)^3 \\ 
&&- 6.555\times 10^{-6}(\Teff)^2 + 0.02245(\Teff) -27.9788.
\end{eqnarray*}
The stellar mass is then estimated via the main-sequence mass-radius relationship
\begin{equation}
\frac{M_*}{M_\odot}\simeq\left(\frac{R_*}{R_\odot}\right)^{1/0.8}.
\end{equation}

\bsp

\label{lastpage}

\end{document}